\documentclass{aa}  

\usepackage{graphicx}
\usepackage{txfonts}
\usepackage{hyperref}
\usepackage{enumitem}
\usepackage{color}
\usepackage{float}

\begin{document}

   \title{Understanding the impact of binary mass transfer in the accretor's measurable parameters}


   \author{M. Vilaxa-Campos
          \inst{1}
          \and
          N. Leigh
          \inst{2} 
          \and
          T. Ryu
          \inst{3, 4, 5}
          }

   \institute{Department of Astrophysics/IMAPP, Radboud University,
   			 PO Box 9010, 6500 GL, The Netherlands \\
   			 \email{magdalena.vilaxa-campos@ru.nl}
		 \and   			  
   			  Departamento de Astronomía, Universidad de Concepción,
              Concepción, Bío Bío\\
              \email{nleigh@amnh.org}
         \and
             Max Planck Institute for Astrophysics,
             Karl-Schwarzschild-Strasse 1, 85748 Garching
         \and
         	JILA, University of Colorado and National Institute of Standards and Technology,
         	440 UCB, Boulder, 80308 CO, USA
         \and
         	Department of Astrophysical and Planetary Sciences,
         	University of Colorado, 391 UCB, Boulder, CO 80309-0391, USA\\
             \email{taeho.ryu@colorado.edu}
             }

   \date{}

 
  \abstract
   {Stars in binaries and higher order systems can experience mass transfer events between their components. The angular momentum carried by the mass gained by the accretor can change the observable parameters of the star and spin it up to critical rotation. In the case of disk accretion, a mass gain lesser than a 10\%  of the accretor’s initial mass is expected to bring it to a critical rotation rate and break it apart.}
   {In this work, we aim to explore the spin-up effect of direct accretion through a stream as a possible mechanism for an accretor to gain more than a tenth of its initial mass without acquiring enough momentum to reach critical rotation.}
   {We present a novel analytical model to characterize the effects of direct mass transfer on the accretor’s measurable parameters as a function of the binary’s semi-major axis and eccentricity and the donor’s rotation velocity. This model takes a two-body approach to the problem, where a stream is decomposed as many discrete particles that do not interact with each other and are influenced by the accretor's gravitational potential only. Each parcel has an instant orbital solution derived from its initial conditions. The contribution each accreted parcel has to the total spin-up of the accretor is given by its tangential velocity at impact, through conservation of angular momentum.}
   {Direct mass transfer proves to be inefficient at spinning up the accretor and thus enables the star to gain a great fraction of its initial mass without reaching critical rotation. We also quantify the fraction of mass that directly impacts the accretor in contrast to the mass that is either lost from the system or creates a disk around a star. Our results show that systems are the most mass conservative when the orbit is tighter or when the donor's spin is greater. In terms of eccentricity, the conservation of mass shows mixed results depending on the system's other initial properties. However, systems with higher eccentricity are consistently a hundred percent conservative within our parameter space.}
   {}


   \maketitle
%

\section{Introduction}

Stars do not always live in isolation. About 85\% of all stars have at least one companion \citep{2023ASPC..534..275O, 2012Sci...337..444S, 2013ASPC..470..141S} and, in particular, roughly half of solar type stars are part of binaries or higher order multiple star systems \citep{Raghavan2010}. 

Throughout their lives, binaries and higher order systems might experience mass transfer events between their components \citep{1971ARA&A...9..183P}, typically referred to as the accretor and donor. Mass transfer can occur through different mechanisms depending on the binary's conditions. For example, closer systems might accrete through wind Roche lobe overflow while systems with wider orbits accrete through Bondi-Hoyle (BH) accretion \citep{2017MNRAS.468.4465C}.

Mass transfer events can affect the orbital parameters of the system \citep{2007ApJ...667.1170S, 2009ApJ...702.1387S, 2010ApJ...724..546S, 1966ARA&A...4...35H, 1977A&A....54..539K} as well as the composition \citep{1984PASP...96..117M} and observational properties (i.e. color, brightness, rotational velocity) of the accretor and the donor \citep{2013ApJ...764..166D}. Blue Stragglers (BSs), for example, are stars that, if members of a cluster, are found in the region brighter and bluer than the main-sequence turn-off (TO) in the  color-magnitude diagram (CMD). These stars were first discovered by \citet{Sandage53} and are thought to be the product of mass transfer events \citep{MathieuGeller, PortegiesZLeigh19, 2020MNRAS.496.1819L}.

The changes to an accreting star's observational properties due to the increase in mass are yet to be fully understood. In regard to the spin-up effect a star suffers under mass gain, \citet{Packet81} estimated that only 10\% of a star's initial mass would need to be accreted by it to spin it up to its critical rotation rate (i.e. the critical angular velocity at which the star reaches centrifugal limit at the equator and can no longer hold itself together) therefore breaking up the star at its outer layers. This would make most accretion products rapid rotators. Rapidly rotating stars are rare, but some have recently been identified in globular clusters via their blue straggler populations. Specifically, rapid rotators seem to prefer low-density cluster environments \citep{Ferraro23} where wider systems have a higher survival rate due to the lower frequency of strong dissociative dynamical interactions between systems.

Previous studies have modeled direct accretion between components of a binary using different methods. For example, \citet{2010ApJ...724..546S} approached modeling direct accretion through three-body integration. In their approach, the donor star looses a small discrete amount of mass strictly at periastron passage, resulting in either self-accretion, direct impact on the accretor or possibly disk formation around the accretor. Their results focus on how the mass transfer event affects the orbital parameters as well as the donor's rotation rate with respect to the system's mean motion. In their study, \citet{2016ApJ...825...70D} (and the following study \citet{2016ApJ...825...71D}) mass transfer is treated as a perturbation to the donor-accretor two-body system. Their results also center on the evolution of the binary orbital elements, but in eccentric systems.

In this work we present an analytical model for direct accretion, with a focus on the accretor's response to mass gain. We reduce the system to an accretor-particle two-body problem in section \ref{Model}, where we also address the impact of neglecting the influence of the donor's potential (section \ref{Validation}). We measure the contribution to the thermal and rotational energy of the accretor caused by the mass gain, as well as the spin-up effect suffered by the accretor after one orbital period (section \ref{rot_and_ther}). We also measure the efficiency of mass transfer, for a range of different semi-major axis lengths, eccentricities and donor rotation rates (section \ref{conservativeness}). Finally, we discuss the timescales for orbital evolution and how they compare with our accretion timescales in section \ref{Discussion}. We summarize our findings in section \ref{Conclusions}.

\section{The Model} \label{Model}

%

We consider a binary star where $m_{\text{don}}$ and $m_{\text{acc}}$ are the masses of the primary (or donor) and secondary (or accretor). The accretor is a main sequence (MS) star and the donor has already evolved to a giant stage. The frame of reference is fixed to the accretor star so that the donor star traces an elliptical orbit around it. We denote by "$a$" the initial value of the binary's semi-major axis, and its eccentricity by "$e$".

The distance from the origin to the L1 point is so that, at that position, the gravitational effects of both the donor and accretor, as well as the centripetal force inflicted on an imaginary particle of velocity $v_i$ are in equilibrium. This condition is met when equation \ref{eq:L1} is fulfilled.
\begin{align} \label{eq:L1}
    \frac{G m_{\text{don}}}{(a - r_{\text{L1}})^2} - \frac{G m_{\text{acc}}}{r_{\text{L1}}^2} + \frac{v_i^2}{r_{\text{L1}}} = 0
\end{align}

The L1 point moves as the donor orbits around the accretor. At every position along L1's trajectory, we imagine that a single parcel of mass is released from the donor's envelope. For this approach, the following assumptions are made:
\begin{enumerate}[label=\Roman*]
    \item The infalling parcel of mass is solely affected by the gravitational force the accretor star inflicts on it. The gravitational potential of the donor is not taken into account once a parcel is released. \label{I}
    
    \item The effect of a stream of mass directly impacting the accretor is approximated by the superposition of every accreted parcel over a single orbital period of the system. \label{II}

    \item Only the tangential component of a parcel's velocity at the moment of impact will contribute to the rotational energy of the accretor. \label{III}

    \item The radial component will instead contribute to the star's thermal energy. \label{IV}
\end{enumerate}
It is important to highlight that this is a rough first-order approximation that we make in order to be able to derive analytic solutions. We quantify the validity of this assumption later in Section \ref{Validation} by comparing to a known solution to this problem.

We take 1000 values for the true anomaly, with a constant time-step between them, over a single orbital period. The donor's orbit is computed from its orbital elements, so both the trajectory of the donor's center of mass and the distance to L1 are independent of the time step size, therefore avoiding time-step-related errors. At each position on the path traced by L1, a parcel is let go from the donor's envelope. Given our second assumption (\ref{II}), we can now focus our analysis on a single parcel.

\begin{figure}
    \centering
    \includegraphics[width=\hsize]{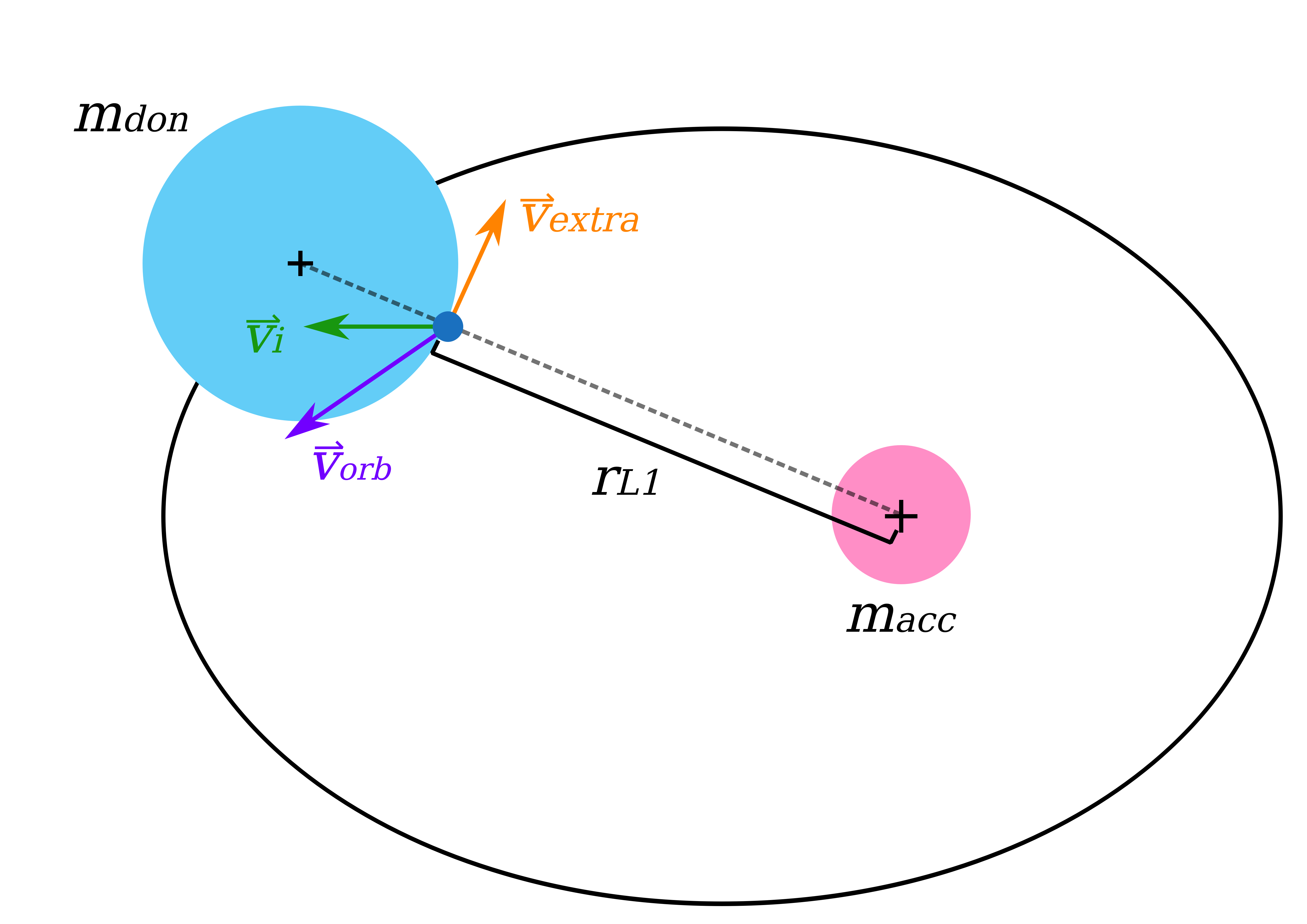}
    \caption{Diagram of the system. The accreting star's position is fixed to the origin of the reference frame while the donor star, of mass $m_{\text{don}}$, orbits around it. A parcel of mass, located at a distance $r_{\text{L1}}$ from the accretor's center of mass, is let go with an initial velocity $\Vec{v}_i = \Vec{v}_{\text{orb}} + \Vec{v}_{\text{extra}}$.}
    \label{fig:sys}
\end{figure}

A parcel drops from a distance $r_{\text{L1}}$ to the accretor's center, with an initial velocity $\Vec{v}_i$. This velocity is calculated as the instantaneous orbital velocity of the donor star's center of mass $\Vec{v}_{\text{orb}}$ in addition to an extra velocity $\Vec{v}_{\text{extra}}$ perpendicular to the radial direction, as depicted in Figure \ref{fig:sys}. The term $\Vec{v}_{\text{extra}}$ acts as the donor's surface linear velocity, given an angular rotation velocity $\omega_{\text{don,rot}}$. In order for $\Vec{v}_{\text{extra}}$ to act opposite to the orbital motion at periastron, it is assumed that $\omega_{\text{don,rot}}$ and the angular orbital velocity of the donor around the accretor $\omega_{\text{orb}}$ have the same direction. For eccentricities greater than 0, $\Vec{v}_{\text{orb}}$ and $\Vec{v}_{\text{extra}}$ are only parallel at periapsis and apoapsis. Any other position in the orbit produces an initial velocity with an angle $\alpha_i$, different from the orbital velocity's angle $\alpha_{\text{orb}}$. A parcel's initial speed $v_i$ and its corresponding angle $\alpha_i$ are given respectively by 
\begin{align} \label{eq:vi}
    v_i = \sqrt{v_{\text{orb}}^2 + v_{\text{extra}}^2 - 2 |v_{\text{orb}}| |v_{\text{extra}}| \cos{\left( \frac{\pi}{2} - \alpha_{\text{orb}} \right)}}
\end{align}
and
\begin{align}
    \alpha_i = \alpha_{\text{orb}} + \arcsin{\left[ \frac{v_{\text{extra}}}{v_i} \sin{\left( \frac{\pi}{2} - \alpha_{\text{orb}} \right)} \right]}
\end{align}
We defer this demonstration to Appendix \ref{apx:initial_angle}.

\begin{figure}
    \centering
    \includegraphics[width=\hsize]{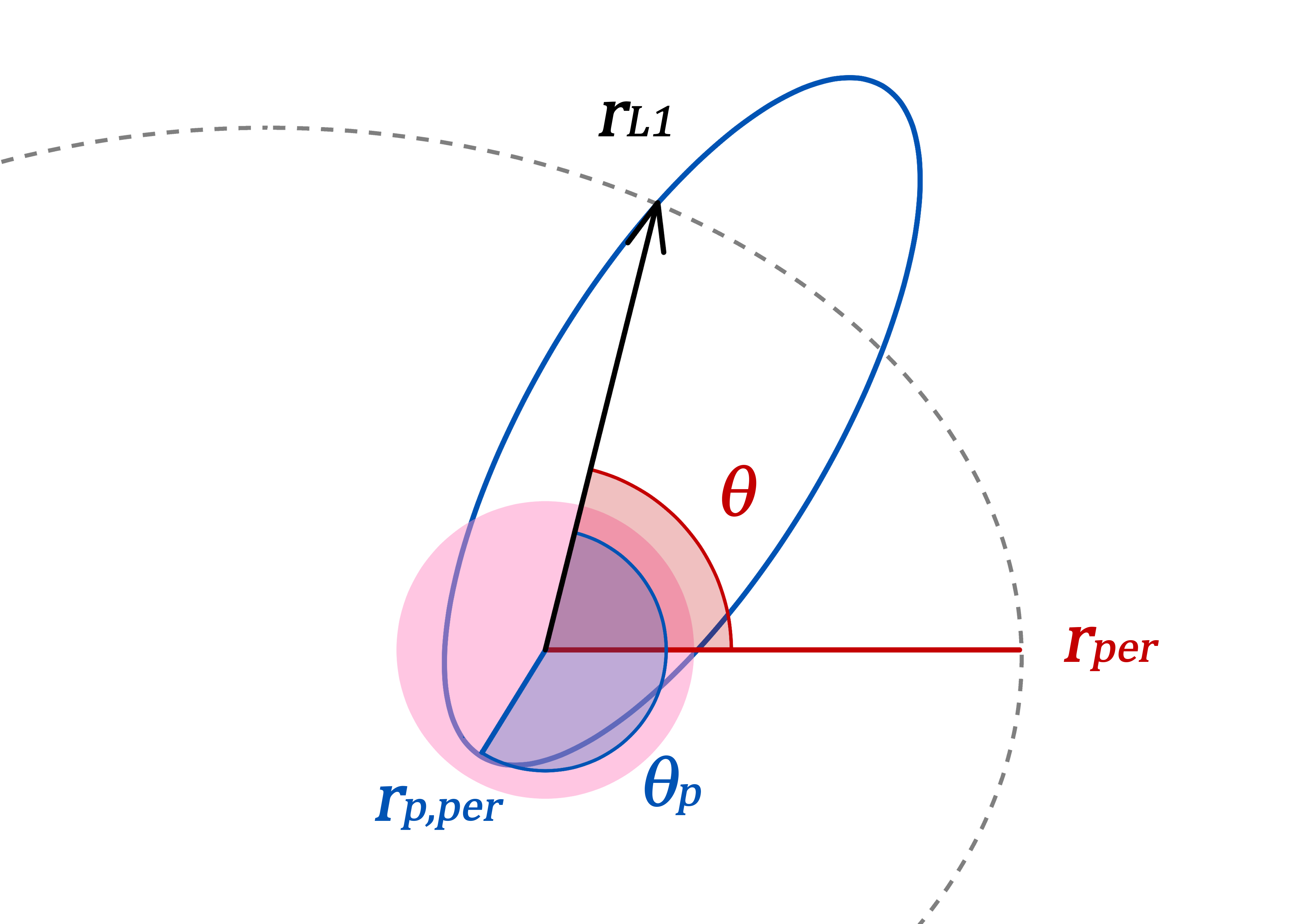}
    \caption{Inclination of a parcel's orbit (blue) with respect to the donor's orbit (in the same direction of the gray dashed $L1$ position). Angle $\theta$ in red represents the true anomaly of the starting point of a parcel with respect to the donor's orbit while $\theta_p$ in blue represents the true anomaly of the parcel's starting point with respect of the parcel's orbit.}
    \label{fig:truean}
\end{figure}

Having a value for the initial velocity and its angle with respect to the radial direction, a trajectory can be predicted for the parcel if we assume it will be in a Keplerian orbit around the accretor. The semi-major axis $a_p$ of this orbit can be obtained through the Vis-viva equation:
\begin{align} \label{eq:visviva}
    v_i = \sqrt{G m_{\text{acc}} \left( \frac{2}{r_{\text{L1}}} - \frac{1}{a_p} \right)}
\end{align}
By solving for the semi-major axis, we get
\begin{align}
    a_p = \left( \frac{2}{r_{\text{L1}}} - \frac{v_i^2}{G m_{\text{acc}}} \right) ^{-1}
\end{align}
The use of the Vis-viva equation to relate the particle's velocity to the semi-major axis of its orbit ensures the conservation of its mechanical energy. The eccentricity $e_p$ can be obtained through 
\begin{align}
    e_p = \frac{c_p}{a_p}
\end{align}
, where
\begin{align} \label{eq:cp}
    c_p = \frac{1}{2} \sqrt{{r_{\text{L1}}}^2 + (2a_p - r_{\text{L1}})^2 - 2r(2a_p - r_{\text{L1}}) \cos{(\pi - 2\alpha_i)}} 
\end{align}
This approach is consistent with the conservation of angular momentum. Assuming the accretor is at one of the foci of the parcel's orbit, the true anomaly of the initial position of the parcel with respect to the new orbit is given by
\begin{align} \label{eq:truean}
    \theta_p (t_p = 0) = \pi - \arcsin{\left[ \frac{2a_p - r}{2c_p} \sin{(\pi - 2\alpha_i)} \right]}.
\end{align}
, where it is important to note that the argument $t_p = 0$ denotes the initial position within the parcel's orbit, and it is not an indicator of the time transpired since the beginning of the donor's orbit at periapsis. Equation \ref{eq:truean}, in combination with $\theta$, the true anomaly of the starting point of the parcel in the L1 orbit, gives us the angle offset of the parcel's orbit with respect to the donor's orbit, as shown in Figure \ref{fig:truean}.

\begin{figure}
    \centering
    \includegraphics[width=\hsize]{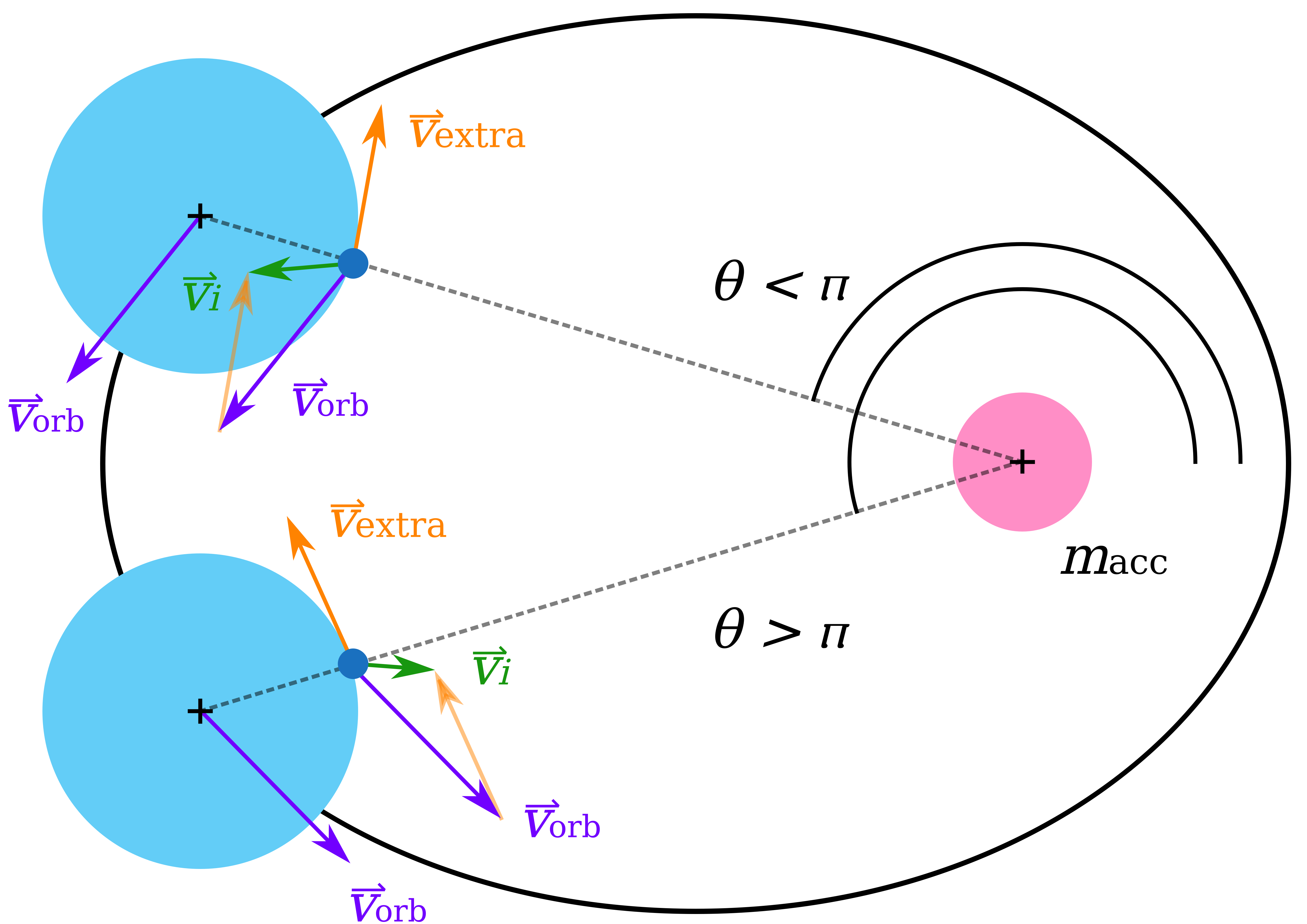}
    \caption{Resulting velocity $\Vec{v_i}$ for a parcel in the first half ($\theta < \pi$) and the second half of the donor's orbit ($\theta > \pi$). The accretor is represented as a pink circle to the right while two different positions of the donor are displayed to the top and bottom left of the figure in light blue. A parcel at the surface of the donor is drawn as a small blue circle. Top left: The angle between $\Vec{v}_{\text{orb}}$ and $\Vec{v}_{\text{extra}}$ is always greater than $\pi$. This produces an initial velocity $\Vec{v}_i$ that always points back towards the donor star ($\alpha_i > \alpha_{\text{orb}}$). Bottom left: The angle between $\Vec{v}_{\text{orb}}$ and $\Vec{v}_{\text{extra}}$ is always smaller than $\pi$. This produces an initial velocity $\Vec{v}_i$ that always points away from the donor star ($\alpha_i < \alpha_{\text{orb}}$).}
    \label{fig:diag2-ab}
\end{figure}

A parcel is considered as accreted only if the following conditions are fulfilled:
\begin{enumerate}
    \item At the starting point of a parcel, the donor star must be overfilling its Roche lobe for the system to undergo Roche lobe overflow (RLOF). We define the parameter $f$ as the fraction of the donor's radius that protrudes beyond the L1 point at any point of the orbit. Although this assumes the donor is a rigid body, this value can be interpreted as the opening angle of the Roche surface for a teardrop-shaped donor in a more realistic scenario. In practice, $f$ acts as a weighting factor for the mass distribution between the particles that the donor loses. \label{cond:of}
    
    \item Given the initial velocity $v_i$ and its angle with respect to the radial direction $\alpha_i$, the resulting trajectory must impact the accretor. Impact is defined as the first point in the trajectory at which the distance from the particle to the center of the system is less than or equal to the accretor's radius. This is
    \begin{align}
        r_p \leq r_{\text{acc}}
    \end{align}
    \label{cond:traj}
    
    \item The angle of the parcel's initial velocity must be less than or equal to the angle of the orbital velocity $\alpha_{\text{orb}}$. This discards any parcel whose initial velocity's direction pushes it back towards the donor star. This condition prevents any mass loss from happening in the first half of the binary's period (i.e. from periastron to apoastron), due to the resulting $\Vec{v}_i$ always pointing outwards, as shown in Figure \ref{fig:diag2-ab}. \label{cond:ang}
\end{enumerate}

During a single orbital period of the binary, a flag is assigned to each parcel to indicate if condition \ref{cond:traj} is met or not. In addition to the initial values, the velocity and angle of impact of each accreted parcel are also stored.

This experiment is repeated for many systems with a range of values for $a$, $e$ and the velocity fraction $v_{\text{extra}}/v_{\text{per}}$. The latter describes the magnitude of the linear rotational velocity at the surface of the donor divided by the donor's linear orbital velocity at periastron passage. This way, $v_{\text{extra}}/v_{\text{per}}$ is a measurement of the donor's rotation: when $v_{\text{extra}}/v_{\text{per}}$ = 0 the donor is non-rotating, while $v_{\text{extra}}/v_{\text{per}} = 1$ describes the case where the donor's surface rotation and its orbital velocity at periastron are equal. When $v_{\text{extra}}/v_{\text{per}} = 1$, a parcel dropped at periastron passage will fall to the accretor with an initial speed $v_i = 0.0$.

Since the value of the overfilling fraction $f$ does not affect the trajectory of an in-falling parcel, this value is added to the dataset after all trajectories have been obtained. This allows us to test the accretor's response to different values of $f$ without running the experiment again. The value of this parameter can be obtained for every position in the L1 orbit as described by equation \ref{eq:f}
\begin{align} \label{eq:f}
    f (\theta) = 1 - \frac{r_{\text{orb}}(\theta) - r_{L1}(\theta)}{r_{\text{orb, per}} - r_{\text{L1, per}}} (1 - f_{\text{per}})
\end{align}
, where $f_{\text{per}}$ is the value of $f$ at periastron.

All negative values (i.e. positions where the donor star does not overflow its Roche lobe) are replaced with 0 in the dataset so that they do not contribute to the total gained mass. When this parameter is added to the datasets, condition \ref{cond:ang} is evaluated, and the accretion flag is updated to reflect if all three conditions are fulfilled.

Lastly, taking $\Dot{M} = 10^{-6} M_{\odot}/yr$ (a mass transfer rate on the lower limit of the range \citet{LS2010} worked with) as the mass loss rate of our donor, the mass contained in each parcel of mass can be obtained as follows:
\begin{align} \label{eq:mp}
    m_p (\theta) = \Dot{M} P \frac{f (\theta)}{\sum\limits_{i} f_i}
\end{align}
where $P$ is the orbital period of the binary, and the summation on the denominator is the total addition of all $f$ values over one orbit.

The routine makes use of AMUSE \citep{2018araa.book.....P, 2013CoPhC.184..456P, 2013A&A...557A..84P, 2009NewA...14..369P} to introduce physical units into the calculation.

In section \ref{Results}, the momentum gained by the accretor after one orbital period is evaluated in terms of our three main parameters: semi-major axis "$a$", eccentricity "$e$" and synchronicity "$v_{extra}/v_{per}$". The conservativeness of the accretion is also analyzed as a function of those three parameters.

We set the initial masses of the accretor and donor to 1.00 $M_{\odot}$ and 1.20 $M_{\odot}$, respectively. In regard to the semi-major axis, we work with $a$ between the following values: the lower limit
\begin{align}
    \min (a) = 1.01 * (r_{\text{acc}} + \max (r_{\text{AGB}}))
\end{align}
, which represents the orbital separation of a circular system where the stars' surfaces are almost touching; and the upper limit condition given by Eggleton's equation \citep{1983ApJ...268..368E}
\begin{equation}
	\max (a) = \max (r_{\text{AGB}}) \frac{0.6q^{2/3} + \ln{(1 + q^{1/3})}}{0.49q^{2/3}}
\end{equation}
, where $q$ is the mass ratio $m_{\text{don}} / m_{\text{acc}}$ (in our case, $\min (a) = 1.23$ AU and $\max (a) = 3.10$), and $\max (r_{\text{AGB}})$ is the largest radius of a star of mass $m_{\text{don}}$ during its asymptotic giant branch stage. From SeBa \citep{1996A&A...309..179P, 2012A&A...546A..70T} we get $\max (r_{AGB}) =$ 263 $R_{\odot}$.



For eccentricity, we choose values between 0.00 to 0.95 to cover the widest range possible of closed orbits. Values for $v_{\text{extra}}/v_{\text{per}}$ ranged between 0.80 and 1.00 (see the Appendix \ref{apx:vfr_limit}). Lastly, $f_{per}$ is set to 0.1 for all tests. We expect a change in $f_{per}$ would only scale up or down the total effect the mass gain causes on the accretor without changing the overall behavior of the accretion.

%

\subsection{Validation of Assumptions} \label{Validation}

\subsubsection{3-body problem}

Under the assumption that a parcel of mass is unaffected by the gravitational potential of the donor beyond $L1$, reduces the problem from three bodies to just two: the accretor and the parcel. In this section, we compare our model to a planar restricted circular three-body problem. We chose this specific case of the three-body problem because it is the only one for which analytical equations of motion can be derived. Any other case requires to be solved numerically.

We consider a donor of mass $m_{\text{don}}$, an accretor of mass $m_{\text{acc}}$ and a parcel of negligible mass. The total mass of the stars in this system is taken as unity, so that
\begin{align}
	m_{\text{don}} = \kappa & & \text{and} & & m_{\text{acc}} = 1 - \kappa
\end{align}
, where $\kappa$ is a constant. In addition, the distance $a$ between the stars is set as the length unit, and the time unit is defined so that
\begin{align}
	\frac{P}{2 \pi} = 1
\end{align}

An inertial frame of reference of coordinate axes $\underline{x}$ and $\underline{y}$ is positioned with its origin at the center of mass between both stars. The donor and accretor are placed on the $\underline{x}$ axis at distances $1 - \kappa$ and $\kappa$ respectively. Additionally, a rotating frame of reference of coordinate axes $x$ and $y$ is positioned so that always $x$ aligns with the line connecting the centers of mass of both stars. The frame rotates at a rate
\begin{align}
	\vec{\omega} = \sqrt{\frac{G (m_{\text{acc}} + m_{\text{don}})}{a^3}} \hat{y} = 1 \hat{y}
\end{align}
Equation 5.10 of \citet{3body} describes the equations of motion in such a system. In our terms, these equations become
\begin{align}
	\ddot{x} = \frac{\partial \Omega}{\partial x} + 2 \dot{y} & & \text{and} & & \ddot{y} = \frac{\partial \Omega}{\partial y} - 2 \dot{x}
\end{align}
, where
\begin{align}
	\frac{\partial \Omega}{\partial x} = x + \frac{(\kappa-1)(\kappa+x)}{((\kappa + x)^2 + y^2)^{3/2}} + \frac{\kappa (\kappa + x - 1)}{((\kappa + x - 1)^2 + y^2)^{3/2}}
\end{align}
and
\begin{align}
	\frac{\partial \Omega}{\partial y} = y + \frac{(\kappa-1)y}{((\kappa + x)^2 + y^2)^{3/2}} + \frac{\kappa y}{((\kappa + x - 1)^2 + y^2)^{3/2}}
\end{align}

The initial position of a particle dropped from the L1 point is
\begin{align}
	\vec{r}_p = (r_{\text{L1}} - \kappa) \hat{x}
\end{align}
, where $r_{\text{L1}}$ is the distance between the accretor and the particle, calculated through equation \ref{eq:L1}.

To compare this three-body simulation to our model as an independent check of the validity of our model assumptions, the massless particle must be given an additional initial velocity against the motion of the donor. Since this case describes a circular motion between the stars, the orbital velocity of the donor and the added velocity are always parallel to each other. Then, the initial velocity can be described by equation \ref{eq:vi_in_vfr}, where the donor's rotation is
\begin{align}
	\vec{v}_{\text{don}} = \vec{\omega} \times \vec{r}_{\text{don}} = | \vec{\omega} | (1 - \kappa) \hat{y}
\end{align}
Now, correcting for the rotation of the frame, we obtain
\begin{align}
	\vec{v}_p &= \left( 1 - \frac{v_{\text{extra}}}{v_{\text{orb}}} \right) \vec{v}_{\text{don}} - \vec{\omega} \times \vec{r}_p \nonumber \\
	&= \left[ \left( 1 - \frac{v_{\text{extra}}}{v_{\text{orb}}} \right) (1 - \kappa) - |\vec{r}_p| \right] \hat{y}
\end{align}

\begin{figure*}
    \centering
    \includegraphics[width=0.9\textwidth]{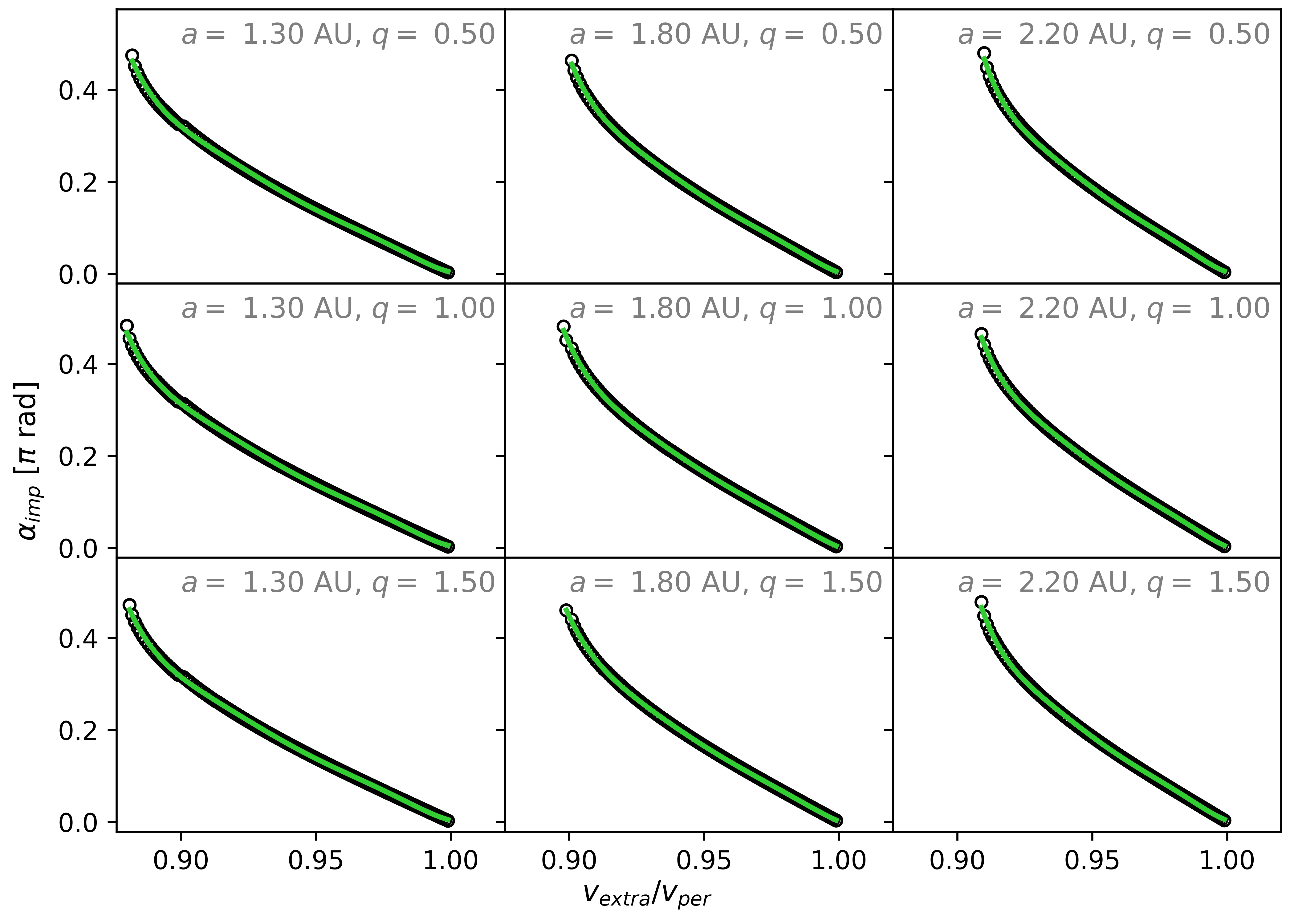}
    \caption{Impact angle $\alpha_{\text{imp}}$ as a function of the donor rotation rate $v_{\text{extra}}/v_{\text{per}}$ in our model. From left to right, the systems have semi-major axes of 1.30, 1.80 \& 2.20 AU. From top to bottom, the systems have mass ratios of 0.50, 1.00 \& 1.50. The data from our model is plotted as open black circles, while the best polynomial fit of order 6 is shown as a continuous green line. The coefficients for the fits can be found in the top half of table \ref{tab:coefficients}.}
    \label{fig:model_fit}
\end{figure*}

\begin{figure*}
    \centering
    \includegraphics[width=0.9\textwidth]{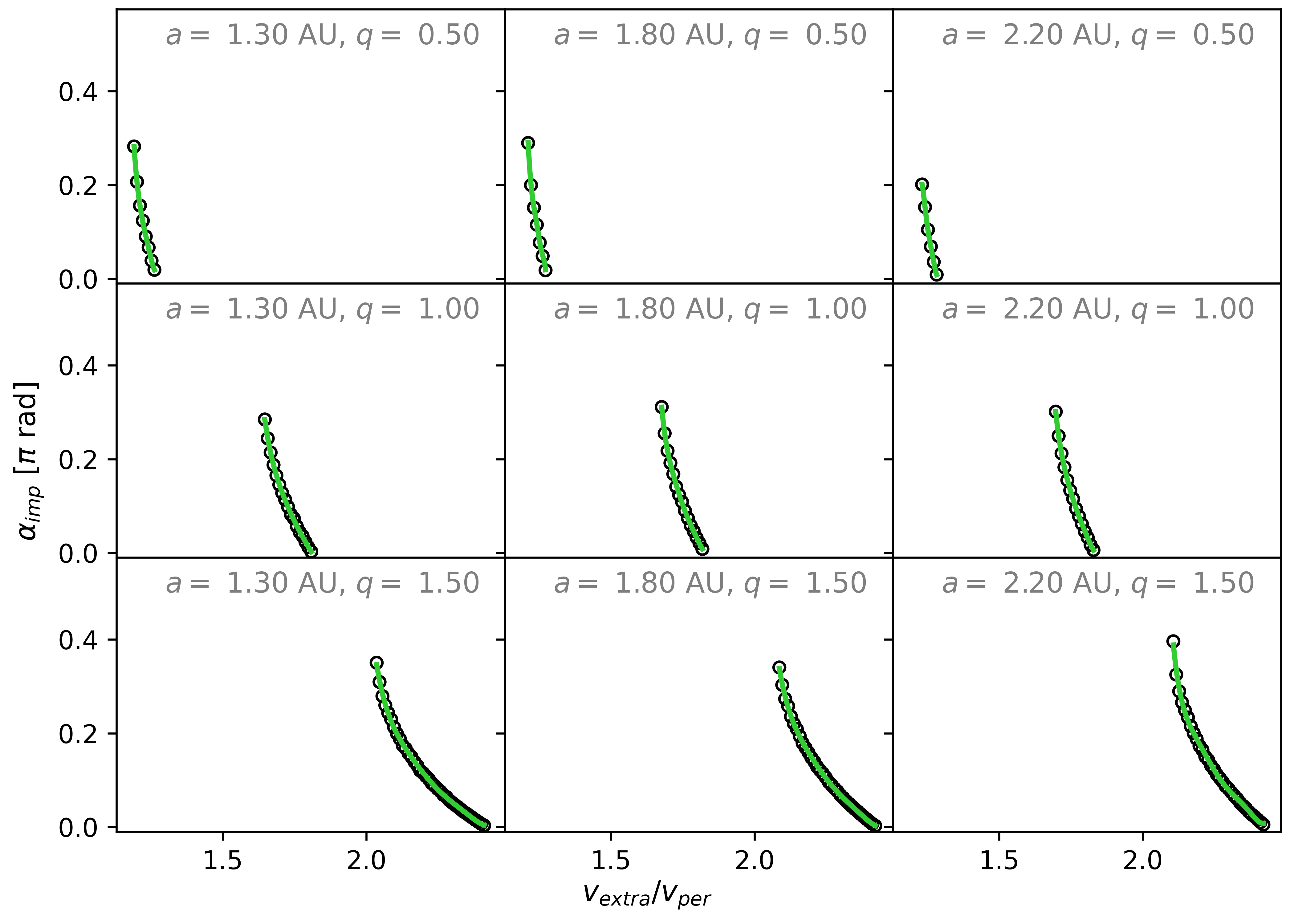}
    \caption{Akin to figure \ref{fig:model_fit} but for the three-body simulations. The coefficients for the fits can be found in the bottom half of table \ref{tab:coefficients}.}
    \label{fig:3b_fit}
\end{figure*}

Having set the initial conditions, the infall is simulated with a time step $dt = 1$ minute. The simulation runs until one of the following three events occurs: i) the particle approaches the accretor at a distance equal or lesser than the accretor's radius, ii) the particle reaches a distance from the accretor greater than the distance to $L1$, or iii) the particle continues to orbit around the accretor for the duration of one orbital period of the system.

Although the impact velocity for both our model and the three-body approach fall within the same order of magnitude, there are two significant differences between the resulting impact angles. First, the three-body approach results are displaced towards higher donor rotation rates in comparison to our model. This is an effect of the donor's gravitational potential being considered in the former. As the donor orbits the accretor, the donor slightly pulls the parcel towards it, resulting in a wider orbit when compared to the same $v_{\text{extra}}/v_{\text{orb}}$ value in our model. In consequence, the donor has to spin faster in the three-body approach than in our model to produce the same results. 

In spite of these differences, the relation between the impact angle and donor rotation rate is well approximated by a polynomial fit of order 6 for both our model and the three-body approach. The best fit for our model and the three-body model are shown in figures \ref{fig:model_fit} and \ref{fig:3b_fit} respectively. The coefficients of the best fit for all systems are displayed in table \ref{tab:coefficients}.

We see that, in our model, $\alpha_{\text{imp}}$ equals 0 when $v_{\text{extra}}/v_{\text{orb}}$ equals 1, regardless of the system's mass ratio and semi-major axis. In contrast, the three-body approach shows a great displacement between systems with different $q$. In both cases, the semi-major axis slightly affects the minimum $v_{\text{extra}}/v_{\text{per}}$ required for direct accretion. We note that since the fits for both models are of the same order, it is possible to find a factor for each polynomial coefficient to make our results align with the results of the three-body approximation. We reserve this analysis for future work.

\begin{figure}[hbt!]
	\centering
    \includegraphics[width=\hsize]{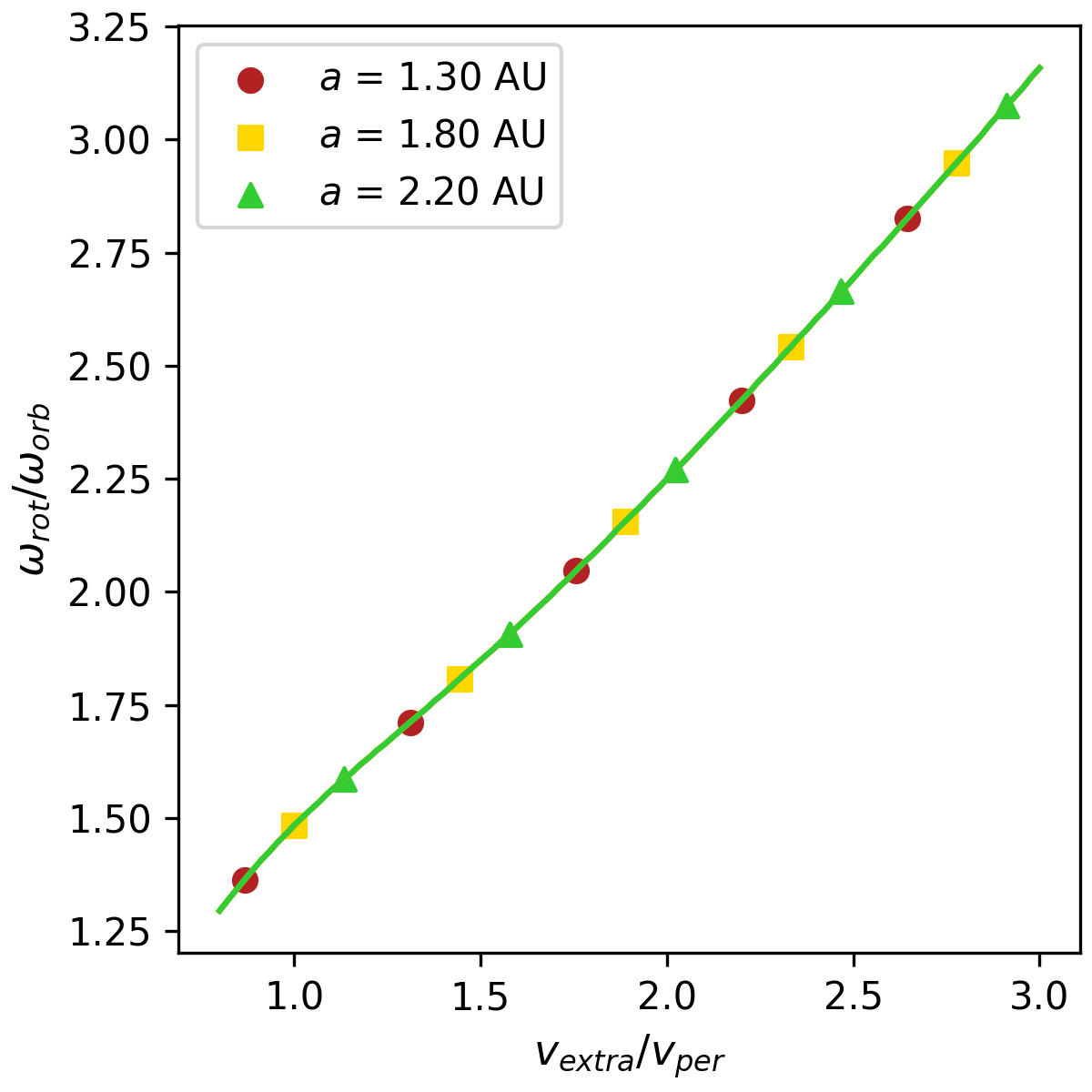}
    \caption{Angular rotational speed of the donor, in terms of the angular orbital velocity of the binary system, as a function of $v_{\text{extra}}/v_{\text{orb}}$. Here, it is assumed that the donor's radius is equal to $a - r_{\text{L1}}$.}
    \label{fig:synch}
\end{figure}

We too note that the range of $v_{\text{extra}}/v_{\text{orb}}$ values explored in this section ($> 0.8$) require the donor's rotation to be supersynchronous to its orbital motion (see figure \ref{fig:synch}). Furthermore, the results from the three-body approach require the donor to fully rotate at least twice for each orbital period. The implications of this will be discussed later in section \ref{Discussion}.

\section{Results} \label{Results}

\subsection{Rotational and Thermal effects} \label{rot_and_ther}

As stated by assumptions \ref{III} and \ref{IV}, the tangential component of the impact velocity of a parcel contributes to the spin-up of the accretor while the radial component adds a thermal contribution. Assuming the accretor behaves as a solid body, the momentum gained from every parcel is given by their mass and velocity at impact with the surface of the accretor. From the impact speed $v_{\text{imp}}$ and impact angle $\alpha_{\text{imp}}$ of each successfully accreted parcel, we can calculate the tangential and radial linear momentum every parcel transfers to the accretor as
\begin{align*}
	p_t = m v_{\text{imp}} \sin{(\alpha_{\text{imp}})} && \& && p_r = m v_{\text{imp}} \cos{(\alpha_{\text{imp}})}
\end{align*}
, respectively. Here, the mass of a parcel is given by equation \ref{eq:mp}. We can then define $\eta$ as the fraction of the total tangential or radial contribution over the total sum of both components, after one orbital period
\begin{align} \label{eq:eta}
	\eta_j = \frac{\sum_i f_i v_{j, i}}{\sum_i f_i (v_{r, i} + v_{t, i})}
\end{align}
Here, $j$ can be either "$r$" or "$t$" to denote the radial or tangential components respectively, and the summations over $i$ go through every parcel that meets all three conditions for direct accretion, over one orbital period.

Since we assume the accretor is a rigid sphere, the accretor's structure is not broken up into layers and the momentum gained by the star is assumed to be distributed uniformly throughout the whole star. We can quantify the amount of momentum transferred to the accretor but, without a more realistic description of the star's internal structure we can not specify how this momentum is transferred as a function of the distance from the accretor's surface (i.e. penetration depth). We comment on this in section \ref{penetrationdepth}.

If we assume the accretor is not rotating before the mass transfer event, its initial rotational angular momentum would be $L_{\text{rot}}=0$. After one orbital period, the accretor would have gained an angular momentum of
\begin{align*}
	\Delta L = \sum_i r_{\text{acc}} p_{t,i}
\end{align*}
, where $r_{\text{acc}}$ is the accretor's radius and the summation over $i$ covers all parcels that directly impact the accretor. From the definition of angular momentum $L = I \omega = I \frac{v}{r}$ and with a moment of inertia $I = (2/5) m_{\text{acc}} r_{\text{acc}}^2$ for a rigid sphere, the accretor would suffer a spin-up of
\begin{align} \label{eq:deltav}
	\Delta v_{\text{rot}} = \frac{5}{2} \frac{\Delta L}{(m_{\text{acc}} + \Delta m) r_{\text{acc}}}
\end{align}
, where $\Delta m$ is the total mass the accretor gained through direct accretion only.

In the following subsections, the proportions of tangential and radial contributions ($\eta_t$ and $\eta_r$), and the spin-up effect suffered by the accretor star after a single orbital period are explored for a range of semi-major axis values, eccentricities and donor rotation velocity values.

\subsubsection{Semi-major axis a} \label{cont_a}

\begin{figure}[hbt!]
	\centering
    \includegraphics[width=\hsize]{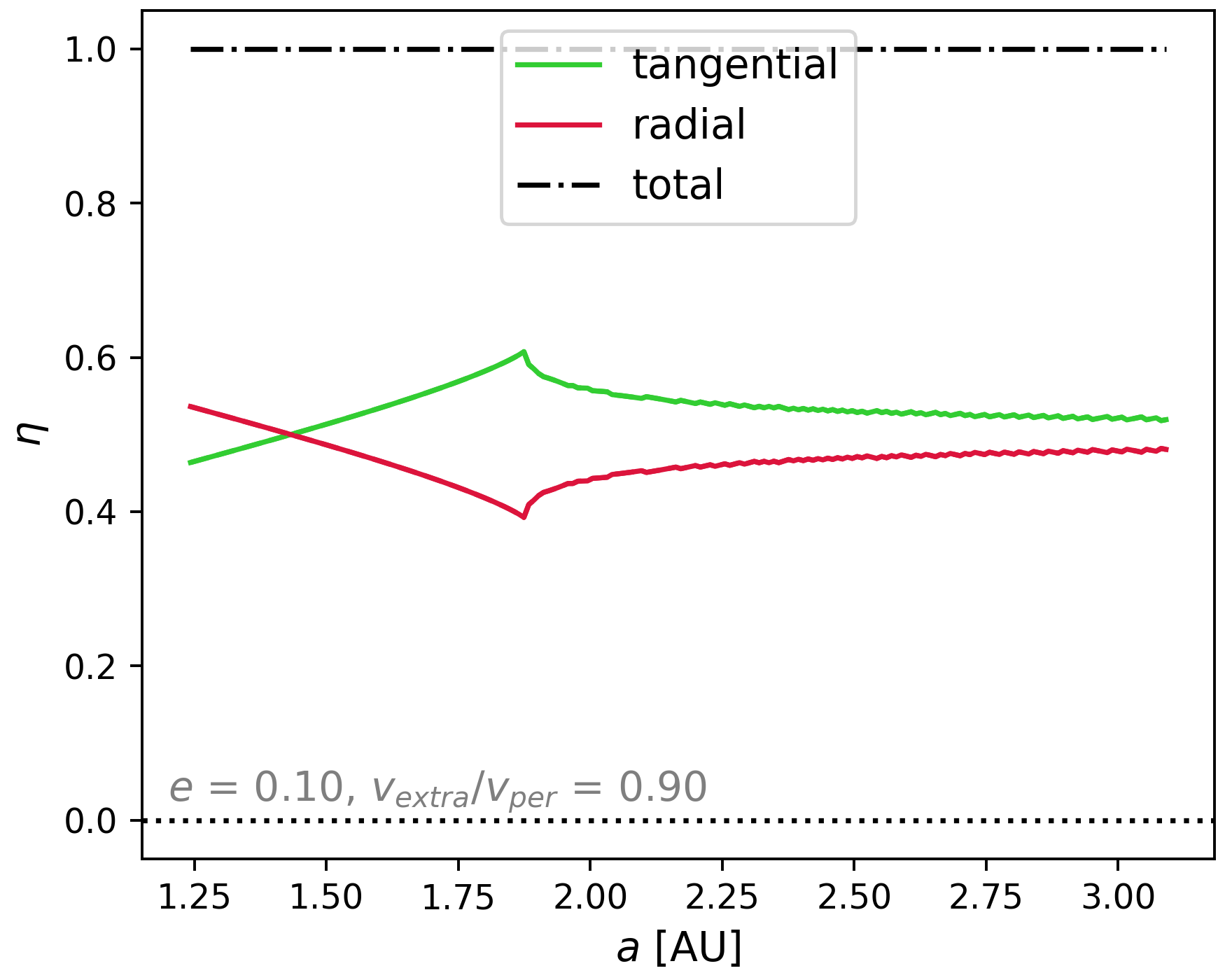}
    \caption{Radial (red) and tangential (green) contributions through direct accretion as a function of the binary system's semi-major axis $a$. The eccentricity is fixed to $e = 0.10$ and the donor's rotation to $v_{\text{extra}}/v_{\text{per}} = 0.90$.}
    \label{fig:vratio_a}
\end{figure}

The value of $\eta$ for both the radial and tangential components, as a function of the semi-major axis, is depicted in figure \ref{fig:vratio_a}.

Closer systems seem to favor the radial contribution over the tangential one. In consequence, we can expect the momentum to be mainly deposited into the accretor's thermal energy rather than its spin for small $a$. Wider systems seem to generally boost the tangential contribution, maximizing the spin-up effect. This is caused by the impact point drifting away from the line connecting the center-of-mass of the accretor to the center-of-mass of the donor, arriving closer to the edge and maximizing the tangential component.

\begin{figure}[hbt!]
	\centering
    \includegraphics[width=0.75\hsize]{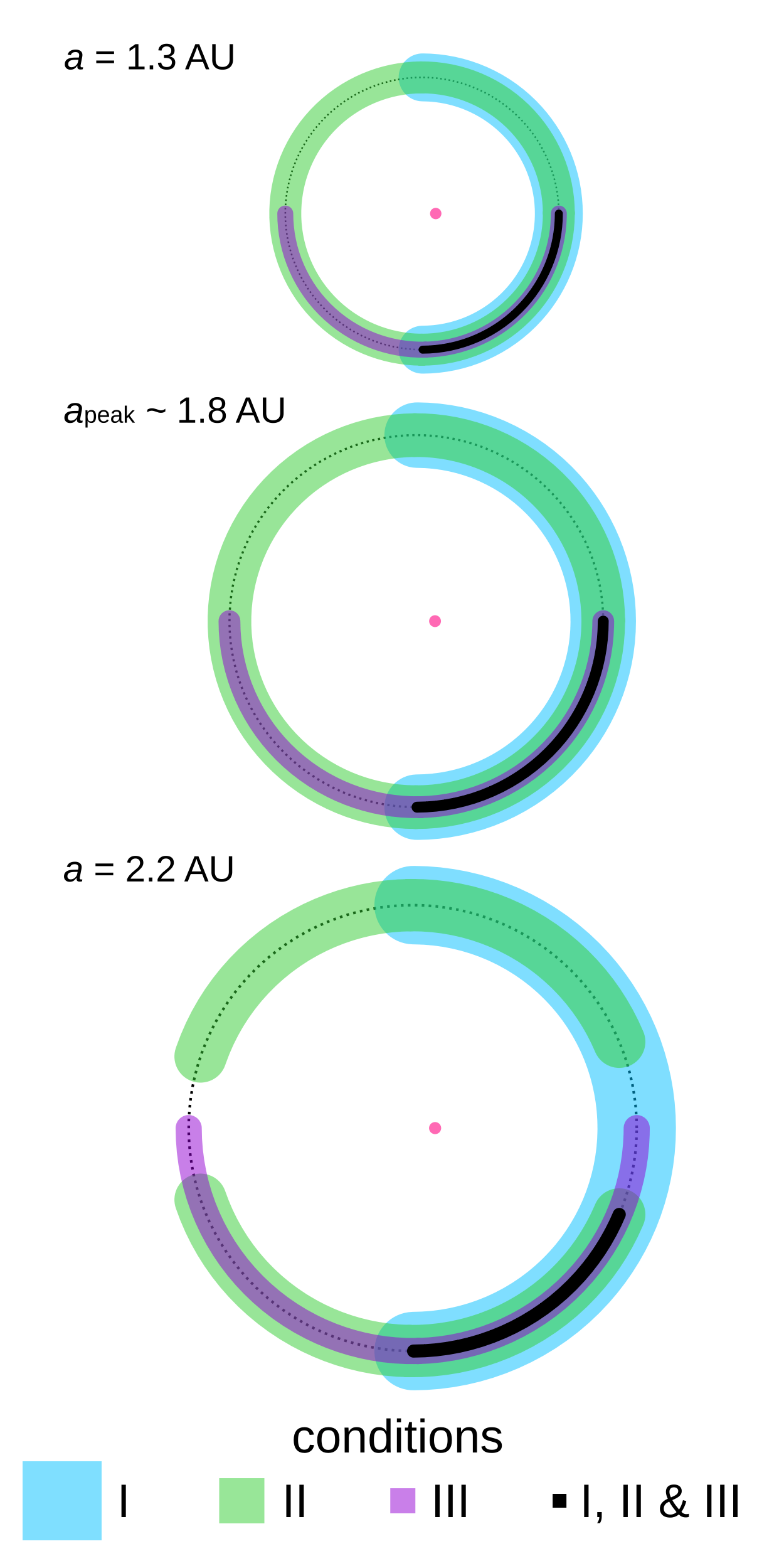}
    \caption{Diagrams showing the region of the donor's orbit where each condition is satisfied. Conditions \ref{cond:of},\ref{cond:traj} and \ref{cond:ang} are colored in blue, green and purple respectively, while the intersection of all three is painted in black. From top to bottom the systems have semi-major axes of 1.30, 1.80 and 2.20 AU.}
    \label{fig:a_diagram}
\end{figure}

There is, however, a drop in the tangential component for $a$ values larger than $\sim 1.86$ AU for a system of eccentricity 0.10 and $v_{\text{extra}}/v_{\text{per}} = 0.90$. This is caused by condition \ref{cond:traj} not being met at periastron. At $a_{\text{peak}}$, the parcel that is dropped from periastron barely touches the surface of the accretor, resulting in a completely tangential impact. For $a \gtrsim a_{\text{peak}}$, the parcel dropped from periastron fails to hit the star. With greater $a$, the section of the orbit where a parcel misses direct impact gradually extends to the parcels around periastron, making the effective direct accretion region (i.e. the region of the orbit where conditions \ref{cond:of}, \ref{cond:traj} and \ref{cond:ang} are met simultaneously) shrink away from periastron while losing the parcels that contribute to the tangential impact the most. The radial component reacts to this by increasing to the right of $a_{\text{peak}}$. Figure \ref{fig:a_diagram} shows a diagram of the regions of the orbit where each condition is met, for different $a$ values.

Although the tangential component decreases and the radial component increases to the right of the peak, the dominance of the tangential component does not invert within our parameter space.

\begin{figure}[hbt!]
    \centering
    \includegraphics[width=\hsize]{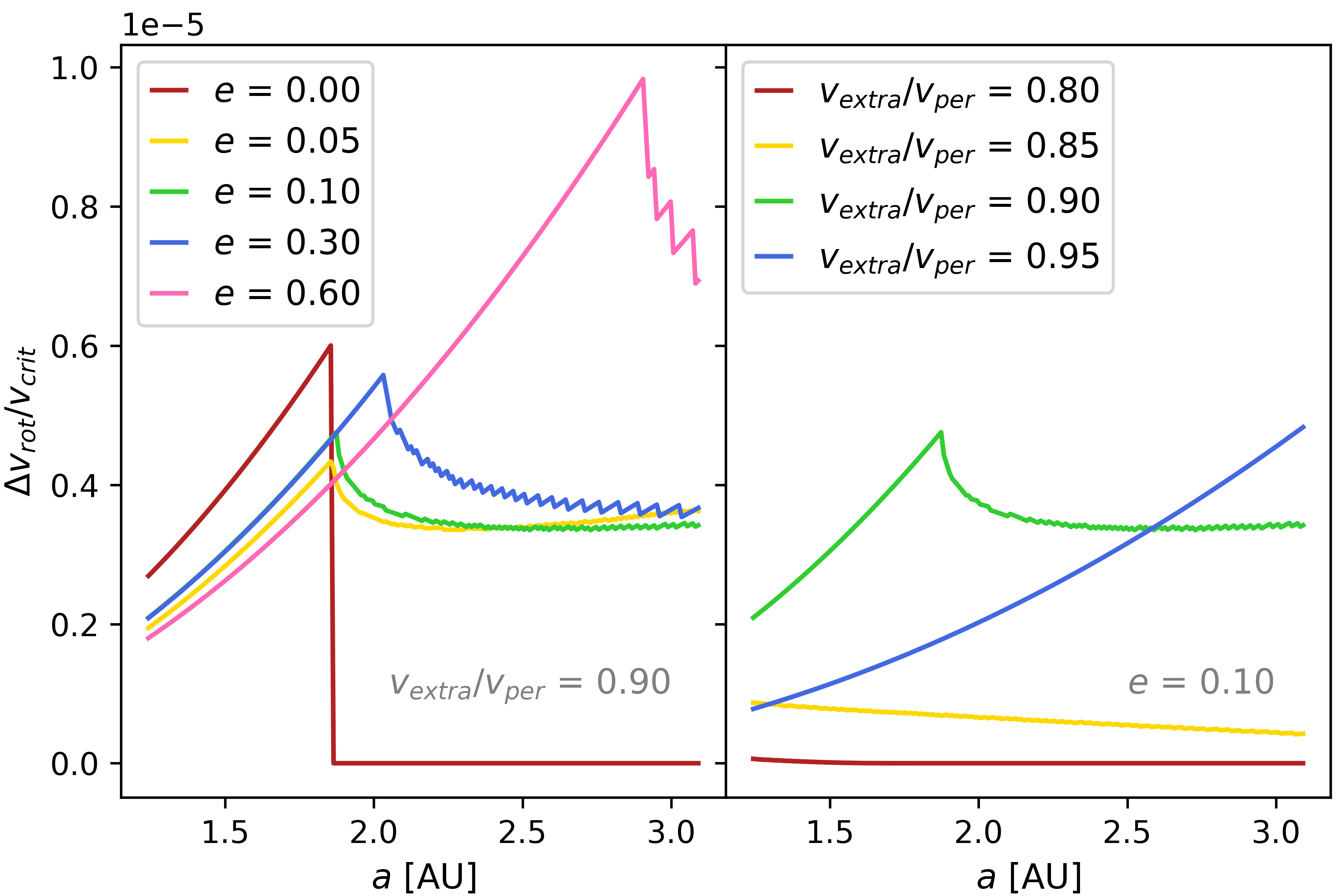}
    \caption{Spin-up of the accretor star in terms of its critical rotation velocity after one orbital period, as a function of the binary system's semi-major axis $a$. Left: All systems have $v_{\text{extra}} / v_{\text{per}} = $ 0.90 and $e$ takes the values 0.00, 0.05, 0.10, 0.30 \& 0.60, shown by different colors (red, yellow, green, blue and pink). Right: All systems have $e = $ 0.1 and $v_{\text{extra}} / v_{\text{per}}$ takes the values 0.80, 0.85, 0.90, \& 0.95 (red, yellow, green and blue).}
    \label{fig:spinup_a}
\end{figure}

The spin-up in rotation an accretor would suffer due to mass gain, through direct accretion, after one orbital period as a function of $a$ is shown in figure \ref{fig:spinup_a}. As expected, the spin-up follows the trend set by the tangential component curve in figure \ref{fig:vratio_a}. In addition to parcels maximizing their tangential velocity for greater $a$, a larger orbital separation means that each parcel will drop from a greater distance, building up speed and impacting the accretor with higher velocities.

In the simplest case of a circular orbit (e.g. red curve on the left panel of figure \ref{fig:spinup_a}), an abrupt break occurs to the right of $a_{\text{peak}}$ instead of the gradual decrease we see in all other curves. A circular orbit makes the initial conditions of every parcel homogeneous. This is, they all have the same initial distance from the accretor surface, the same initial speed, and initial angle. In consequence, this translates to identical impact velocities and angles for every accreted parcel. Then, if for any value of $a$ a parcel fails to impact the star, none of them impact the star, ceasing direct accretion altogether.

\begin{figure}[hbt!]
    \centering
    \includegraphics[width=\hsize]{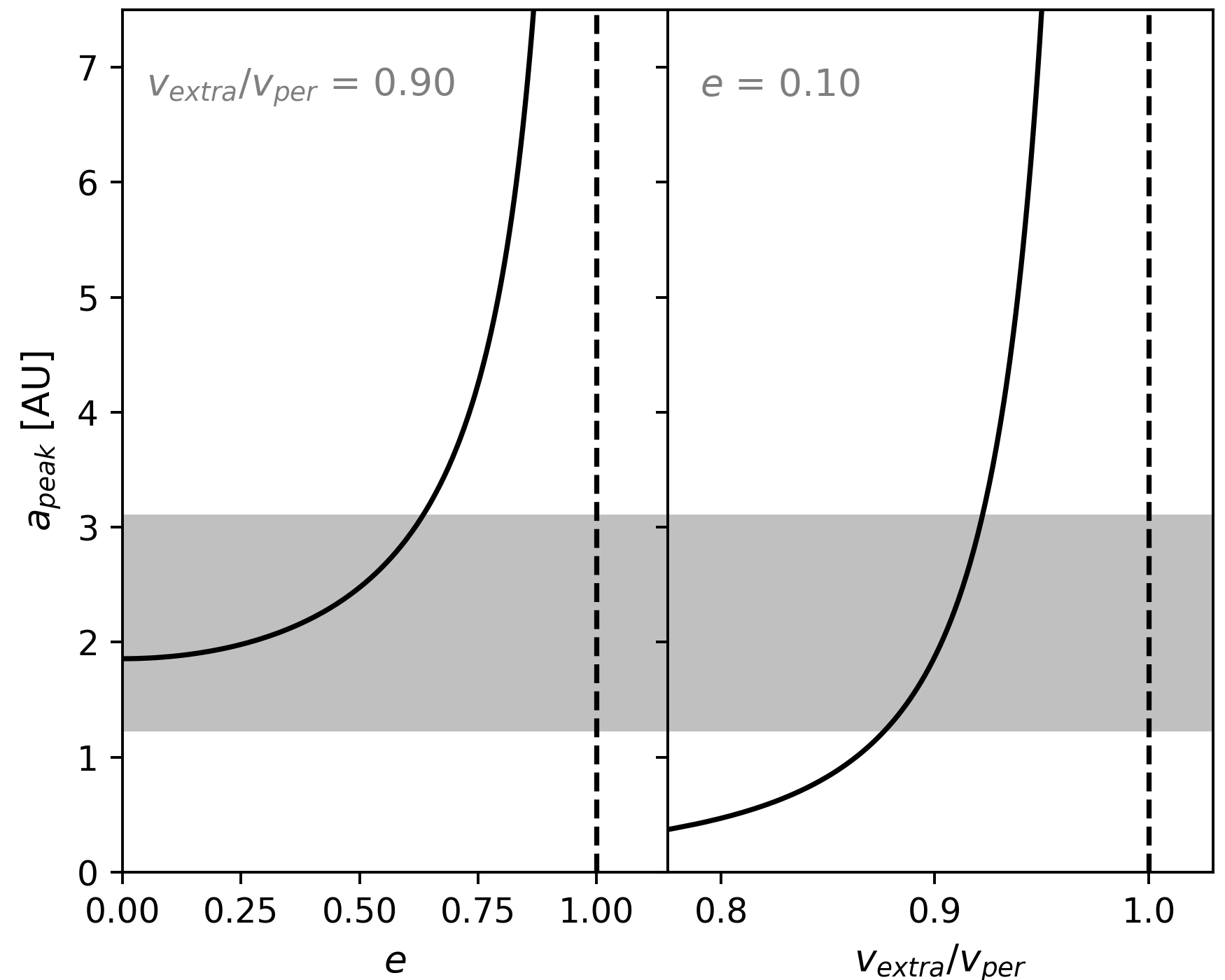}
    \caption{Semi-major axis value for maximum spin-up of the accreting star as a function of the system's eccentricity with constant $v_{\text{extra}} / v_{\text{per}} = 0.90$ (left) and the donor's rotation with constant $e = 0.10$ (right). The shaded region marks the range of $a$ we work with in the analysis.}
    \label{fig:apeak}
\end{figure}

A system with $v_{\text{extra}} / v_{\text{per}} = 0.90$, $e = 0.00$, $m_{\text{acc}} = 1 M_{\odot}$ and $m_{\text{don}} = 1.2 M_{\odot}$ is the most effective at spinning up its accretor when $a = 1.86$ AU. In the general case, the $a$ value for the peak in spin-up can be obtained by following a similar analysis to that of Appendix \ref{apx:vfr_limit}. Having a fixed value for $v_i$, equations \ref{eq:viper} and \ref{eq:vi_in_vfr} can be rearranged into a new expression for the semi-major axis value at which the peak in spin-up occurs:
\begin{align} \label{eq:apeak}
    a_{peak} = \frac{1+e}{1-e} \left( r_{L1} + \frac{r_{L1}^2}{r_{\text{acc}}} \right) \left( 1 - \frac{v_{\text{extra}}}{v_{\text{per}}} \right) ^2 \left( \frac{m_{\text{acc}} + m_{\text{don}}}{2 m_{\text{acc}}}  \right)
\end{align}
By replacing this into equation \ref{eq:L1} for $r = r_{\text{per}} = a (1 - e)$, $r_{L1}$ can be obtained numerically and replaced back into equation \ref{eq:apeak}. The results are shown in figure \ref{fig:apeak}.

On the left panel of figure \ref{fig:apeak}, $a_{\text{peak}}$ is graphed as a function of $e$. The semi-major axis value at which the peak in spin-up occurs grows with eccentricity, showing a clear asymptote at $e=1.0$.
On the right panel, $a_{\text{peak}}$ is shown as a function of $v_{\text{extra}} / v_{\text{per}}$ instead. Within our $a$ range, $a_{peak}$ seems very sensitive to small changes in the donor's rotational velocity, when compared to the more gradual growth portrayed in the left panel. As $v_{\text{extra}} / v_{\text{per}}$ approaches 1, the initial velocity of the parcel at periastron gets closer to $v_i = 0$, making the centripetal force term in equation \ref{eq:L1} negligible, this means the parcel's impact on the accretor's surface will be almost entirely radial and the spin-up provided by it will solely depend on the impact speed which is maximized by increasing the initial distance from which the parcel falls.

The behavior of spin-up for eccentric systems is analogous to the circular case to the left of $a_{\text{peak}}$. To the right of the peak, the curve much closely resembles the gradual drop seen in figure \ref{fig:vratio_a}.

\subsubsection{Eccentricity e} \label{cont_e}

For $e \sim 1$, the spatial resolution at periastron decreases if we choose data points separated by a constant time step $dt$. This can induce resolution-related errors in the calculation of the tangential and radial components, as well as angular momentum, in the vicinity of periapsis. We solve this issue by choosing data points along the orbit that are separated by a constant $d\theta$ instead of setting a constant time step between them. This results in a distribution that does not follow Kepler's law but provides a better resolution for orbits where accretion occurs mostly at periastron. We adhere to our chosen value of 1k data-points per orbit. Increasing this number does not change the results.

\begin{figure}[hbt!]
	    \centering
    \includegraphics[width=\hsize]{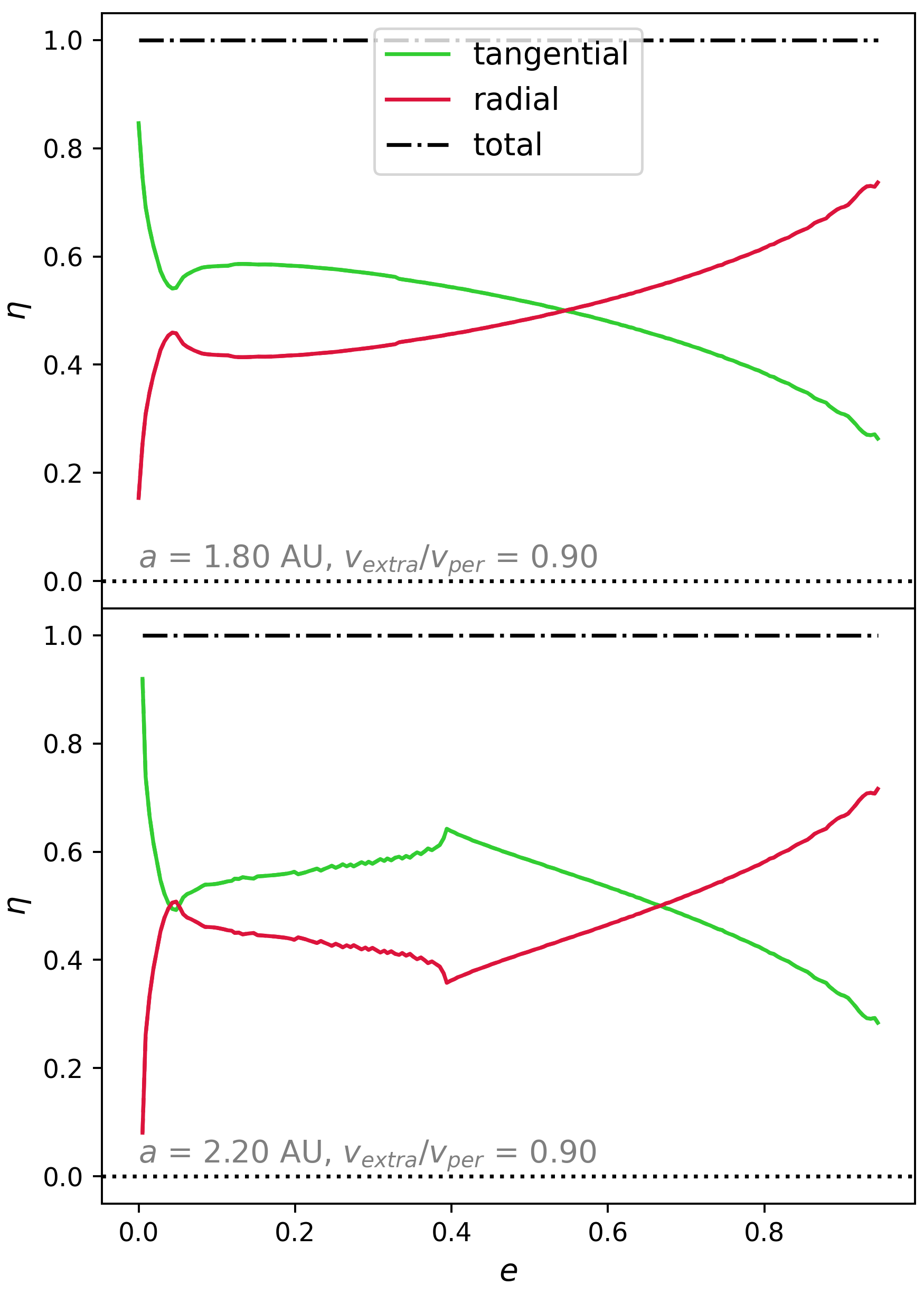}
    \caption{Akin to figure \ref{fig:vratio_a} but as a function of the binary system's eccentricity $e$. Top: The semi-major axis is fixed to $a = 1.80$ AU and the donor's rotation to $v_{\text{extra}}/v_{\text{per}} = 0.90$. Bottom: The semi-major axis is fixed to $a = 2.20$ AU and the donor's rotation to $v_{\text{extra}}/v_{\text{per}} = 0.90$.}
    \label{fig:vratio_e}
\end{figure}

Figure \ref{fig:vratio_e} shows $\eta_r$ and $\eta_t$ as functions of the system's eccentricity, for systems with a donor's rotation value of $v_{\text{extra}}/v_{\text{per}} = 0.90$ and a semi-major axis of 1.80 AU and 2.20 AU in the top and bottom panels respectively. From now on, we will call systems that behave like the top panel of figure \ref{fig:vratio_e} "case 1", and those that resemble the bottom panel "case 2".

\begin{figure*}
    \centering
    \includegraphics[width=0.9\textwidth]{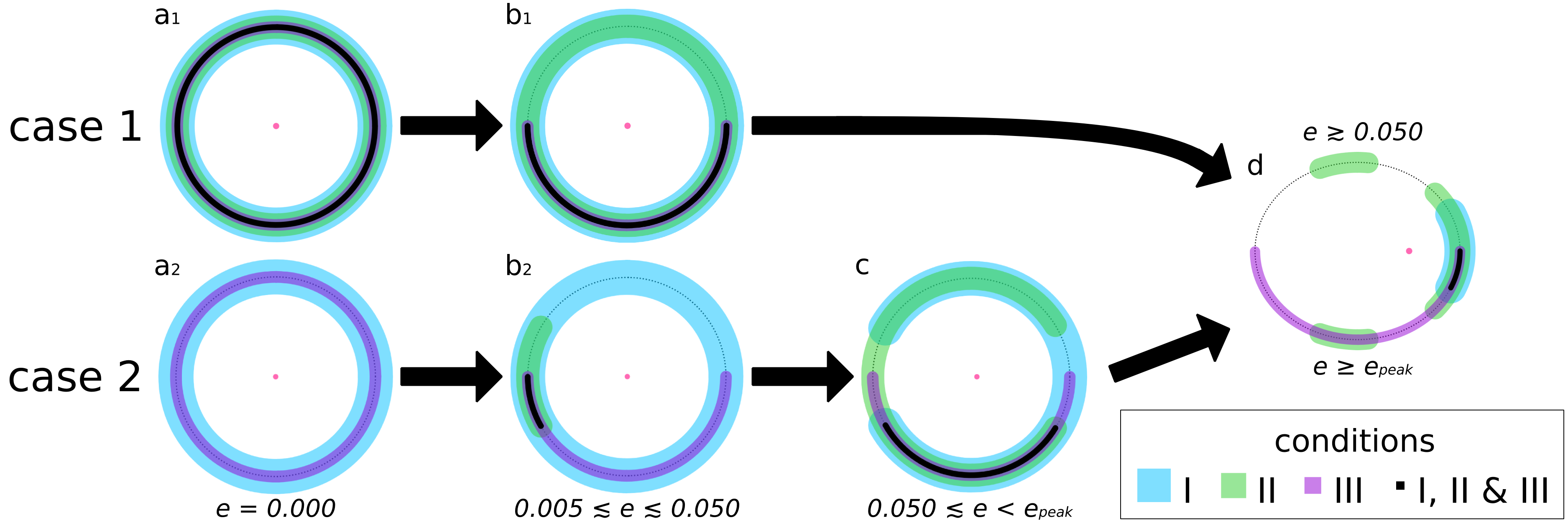}
    \caption{Graphic representation of cases 1 and 2, of the eccentricity spin-up analysis. The accretor star is depicted by a pink circle whose size has been scaled by a factor of 10. A dotted line draws the orbit of the donor's center of mass and, from widest to narrower, the blue, green and purple regions represent areas of the orbit where conditions \ref{cond:of}, \ref{cond:traj} and \ref{cond:ang} are met, respectively, while the black region represents the intersection of all three. Case 1 is exemplified by the yellow curve ($a = 1.80$ AU) on the left panel of figure \ref{fig:spinup_e}, and goes contains stages a$_1$, b$_1$ and d for increasing eccentricities. Case 2 is exemplified by the green curve ($a = 2.20$ AU) on the left panel of figure \ref{fig:spinup_e}, and goes contains stages a$_2$, b$_2$, c and d for increasing eccentricities.}
    \label{fig:ediagram}
\end{figure*}

In case 1, the tangential component dominates on circular orbits. Consequently, direct accretion in systems with $e = 0.00$ is more effective at spinning up the accretor. At $e = 0.00$, all three conditions are satisfied over the entire orbit. For $e \gtrsim 0.00$, condition \ref{cond:ang} reduces the effective accretion area to the latter half of the orbit (i.e. from past apoastron to periastron), causing the drop in the tangential contribution immediately to the right of $e=0.00$. The radial component rises in response. A more subtle decrease/increase in the tangential/radial component occurs towards $e = 0.05$ where a local valley/peak is reached. The slightly greater eccentricity reduces the orbital velocity of the donor at apoapstron, causing the initial velocity of a parcel dropped from apoastron to also decrease. A smaller $v_i$ causes the parcel to fall to the accretor on a more eccentric orbit, impacting its surface in a primarily radial direction, pumping the radial contribution and causing the local peak.

Eccentricities greater than $\sim 0.05$ show a moderate decline/increase in their tangential/radial component. This is caused by conditions \ref{cond:of} and \ref{cond:traj} shrinking the effective region for direct accretion around periastron. At larger eccentricities ($e \gtrsim 0.55$) the radial component takes over, causing most of the momentum from the direct accretion to go into the thermal energy of the accretor. A diagram showing the evolution of a system under case 1 can be found at the top of figure \ref{fig:ediagram}.

Even though it looks like case 2 systems behave identically to case 1 systems for circular orbits, case 2 shows no direct accretion for $e = 0.00$. Condition \ref{cond:traj} is not met anywhere when the orbit is circular, which is coherent with the results shown on the left panel of figure \ref{fig:spinup_a}. For $e > 0.00$, but still close to 0.00, all three conditions for direct accretion are satisfied in a reduced region after apoastron passage. At these small eccentricities, the parcels dropped around apoastron impact the accretor almost completely tangentially.

Greater $e$ causes the effective area for direct accretion to extend from apoastron towards periastron. This expansion of the effective accretion area is, however, countered by the declining tangential contribution from apoastron discussed in case 1's analysis around $e \sim 0.05$, producing a similar drop in the tangential component at the same eccentricity. A distinct feature around $e \sim 0.40$ is present for case 2. The behavior of these curves between eccentricities $0.05 - 0.40$ can be explained by two events occurring simultaneously: the expansion of the effective area for condition \ref{cond:traj} towards periastron and the shrinking of condition \ref{cond:of}'s effective area around periastron. At $e_{\text{peak}}$, the region of the orbit where condition \ref{cond:traj} is fulfilled reaches periastron, where accretion is the most tangential, producing the small peak in the tangential contribution and, accordingly, the local minimum in the radial contribution. For eccentricities greater than that of the peak, case 2 systems behave exactly like case 1 systems. A visual representation of case 2 systems is shown in the bottom row of figure \ref{fig:ediagram}.

\begin{figure}[hbt!]
    \centering
    \includegraphics[width=\hsize]{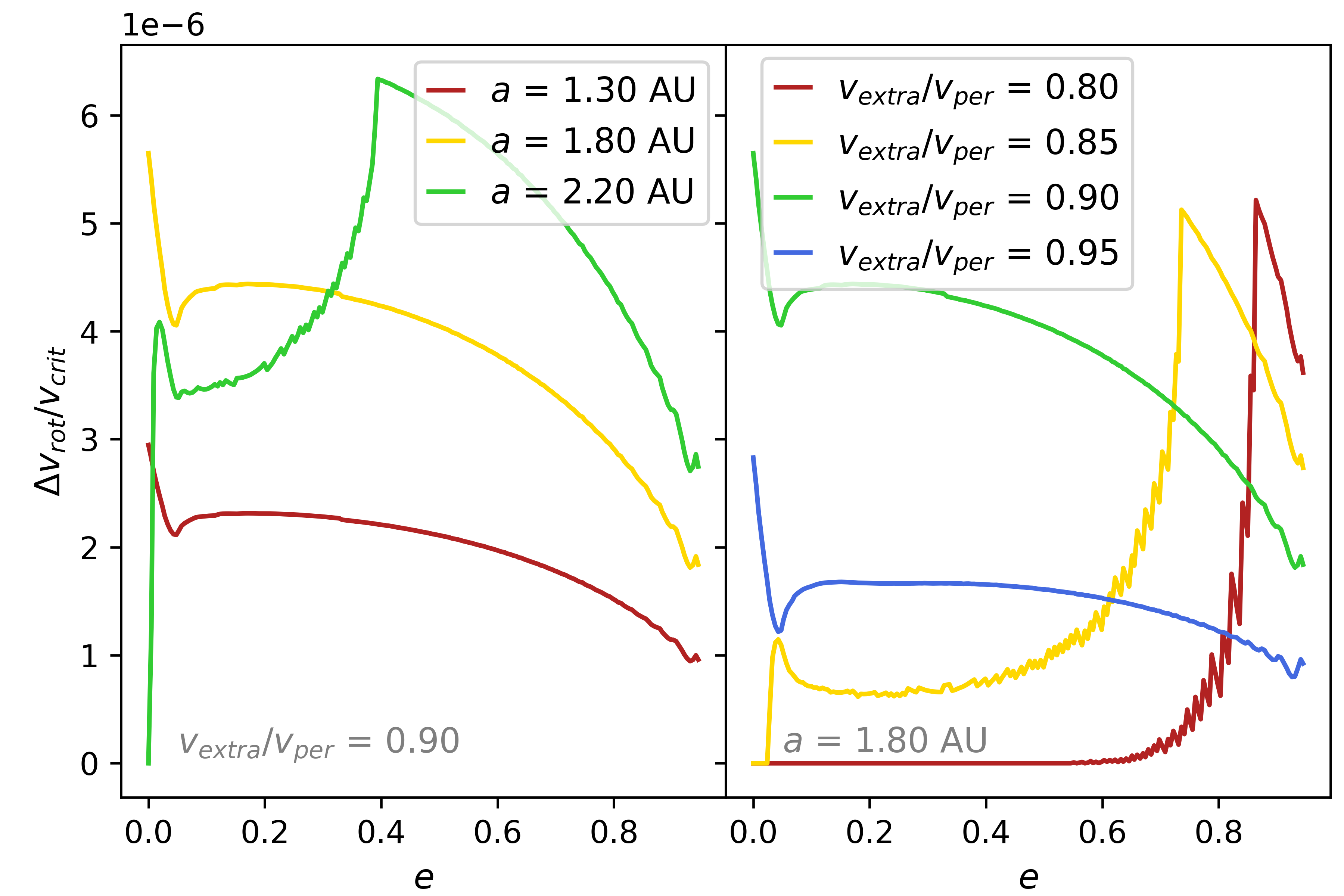}
    \caption{Spin-up of the accretor star in terms of its critical rotational velocity, as a function of the binary system's eccentricity $e$. Left: All systems have $v_{\text{extra}} / v_{\text{per}} = $ 0.90. The result is displayed for $a = $ 1.30, 1.80 \& 2.20 AU (red, yellow and green). Right: All systems have $a =  1.80$ AU and $v_{\text{extra}} / v_{\text{per}}$ takes the values 0.80, 0.85, 0.90, \& 0.95 (red, yellow, green and blue).}
    \label{fig:spinup_e}
\end{figure}

The accretor's spin-up after one orbital period as a function of the system's eccentricity is showcased in figure \ref{fig:spinup_e}.

The spin-up of case 1 systems (e.g. $a = 1.80$ curve on left panel) mimics the behavior of their tangential contribution curves. 

Case 2 systems (e.g. $a = 2.20$ curve on left panel) present some differences between their tangential contribution curve and their spin-up curve, particularly to the left of $e_{\text{peak}}$. As mentioned before, case 2 systems with $e = 0.00$ do not present direct accretion, therefore no spin-up is caused. This explains the abrupt rise in spin-up between circular orbits and $e \gtrsim 0.00$. The feature at $e \sim 0.40$ is also more intense in the spin-up curve. Although the events that explain the spin-up behavior in $0.05 < e < 0.40$ are the same as for the tangential contribution curve, the calculation of the change in spin takes the amount of transferred mass into account. Within $0.05 < e < 0.40$, systems with eccentricities closest to 0.05 mostly accrete from around apoastron where overflow is minimized, thus accreting less massive parcels and gaining less momentum from them, while systems with eccentricities closest to 0.40 mostly accrete from the region around periastron where parcels are the most massive, therefore providing more momentum to the rotation.

It is worth mentioning that case 1 and case 2 correspond to systems with initial conditions to the left and to the right of the $a_{\text{peak}}$ of figure \ref{fig:spinup_a}, respectively.

\subsubsection{Donor's rotation} \label{cont_v}

\begin{figure}[hbt!]
	    \centering
    \includegraphics[width=\hsize]{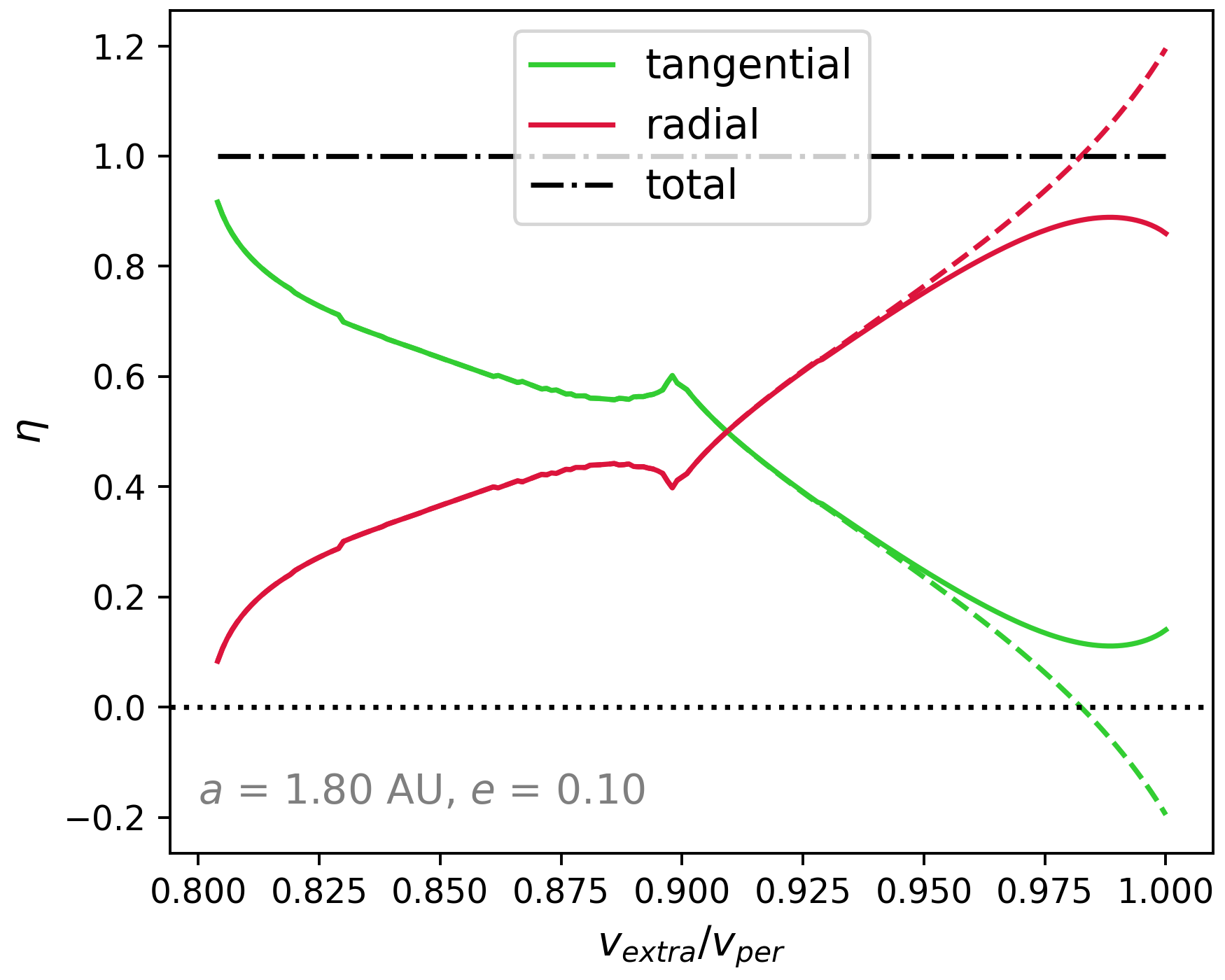}
    \caption{Akin to figure \ref{fig:vratio_a} but as a function of the donor's rotational velocity $v_{\text{extra}} / v_{\text{per}}$. The semi-major axis is fixed to $1.80$ AU and the eccentricity to $e = 0.10$. The solid lines show $\eta$ as described in equation \ref{eq:eta} while the dashed lines show the result of using the absolute value of the tangential velocity instead.}
    \label{fig:vratio_v}
\end{figure}

Unlike figures \ref{fig:vratio_a} and \ref{fig:vratio_e}, to display $\eta$ as a function of $v_{\text{extra}}/v_{\text{per}}$ we make a slight change to equation \ref{eq:eta}. Some combinations of initial conditions produce parcel trajectories that impact the accretor in a retrograde manner. This retrograde velocity introduces a negative factor to some velocities in the summation, producing a negative tangential contribution at $v_{\text{extra}}/v_{\text{per}} \sim 1.00$. Since we want to compare the proportion between tangential and radial contributions, we ignore whether the contribution to spin is prograde or retrograde. To do this, we replace the tangential speed $v_t$ with its absolute value $|v_t|$. The results of $\eta$ as described in \ref{eq:eta}, and with the added absolute value to the tangential speed are shown in figure \ref{fig:vratio_v} as the dashed and solid lines, respectively. This way, we can visualize the proportion between both components in the solid lines while still being able to see where retrograde rotation takes over in the dashed lines.

At the smallest $v_{\text{extra}} / v_{\text{per}}$, the tangential component strongly dominates over the radial contribution, reaching $\eta_t \sim 0.90$. Greater $v_{\text{extra}} / v_{\text{per}}$ provides a reduced tangential contribution and an increased radial contribution. The radial contribution becomes greater than its counterpart for $v_{\text{extra}} / v_{\text{per}} \gtrsim 0.91$. In general, the contribution to thermal energy increases with $v_{\text{extra}} / v_{\text{per}}$, while the spin-up effect is the most effective for smaller $v_{\text{extra}} / v_{\text{per}}$.

\begin{figure}[hbt!]
	\centering
    \includegraphics[width=0.75\hsize]{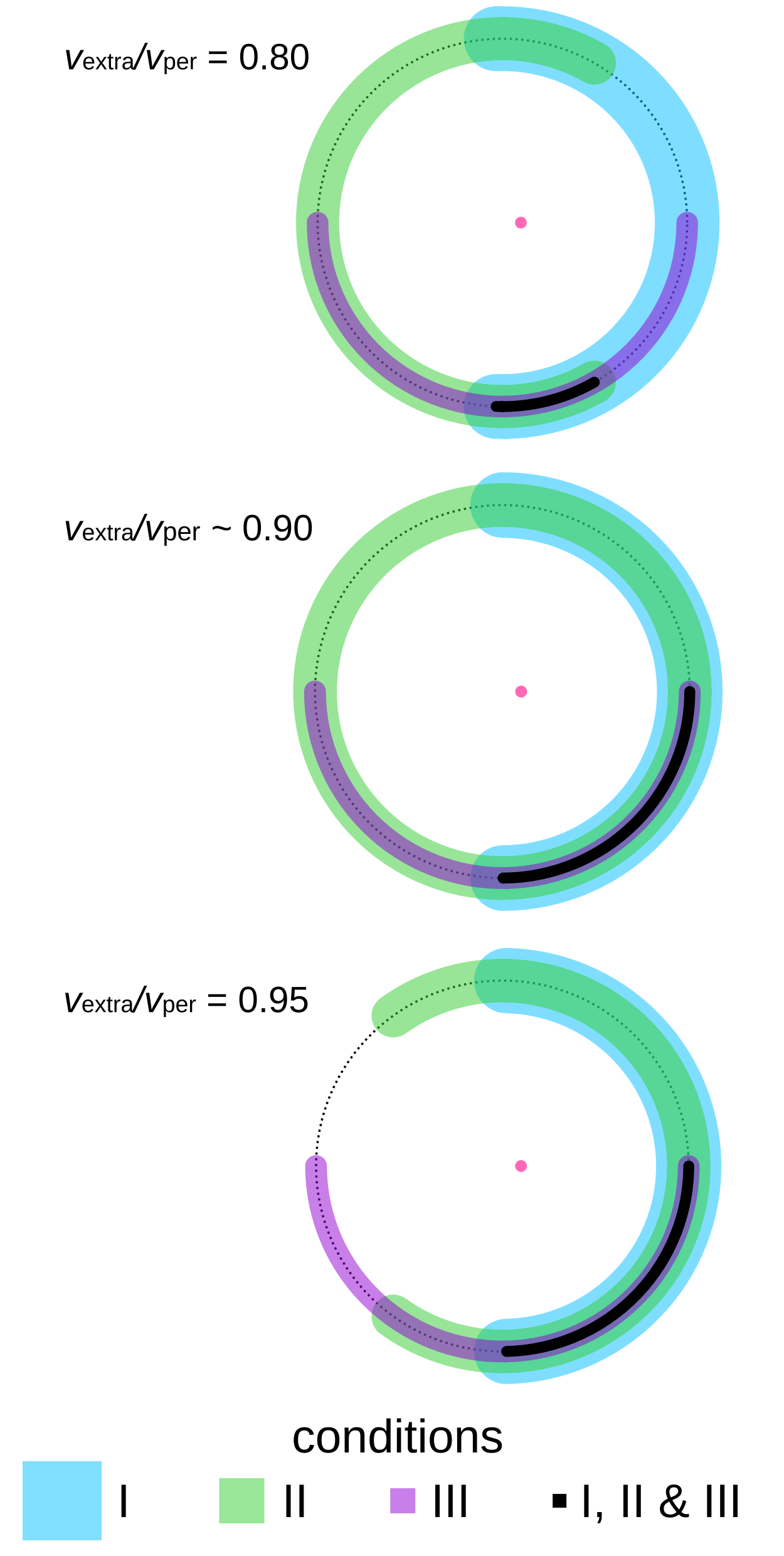}
    \caption{Alike figure \ref{fig:a_diagram}. From top to bottom the systems have $v_{\text{extra}} / v_{\text{per}}$ values of 0.85, 0.90 and 0.95.}
    \label{fig:v_diagram}
\end{figure}

Around $v_{\text{extra}} / v_{\text{per}} \sim 0.90$ a local maximum/minimum is seen in the tangential/radial contribution. The reason behind this feature is similar to that of the one in figure \ref{fig:vratio_a}. In this case, for systems with lower $v_{\text{extra}} / v_{\text{per}}$ values than that of the peak, only a small region of the orbit produces direct accretion. Within the region where the donor overfills its Roche lobe, only the section that is farthest from periastron meets condition \ref{cond:traj}. An increase in $v_{\text{extra}} / v_{\text{per}}$ extends the domain of condition \ref{cond:traj} towards periapsis where the peak in the tangential contribution is reached. At that $v_{\text{extra}} / v_{\text{per}}$ value, the parcel dropped from periastron impacts the accretor completely tangentially. Higher prograde rotation velocities produce more eccentric trajectories for the parcels of mass, impacting the accretor in a mostly radial direction and therefore minimizing the transferred angular momentum. Figure \ref{fig:v_diagram} contains a visual representation of the regions of the donor's orbit where the conditions for direct accretion are met.

At $v_{\text{extra}} / v_{\text{per}} \gtrsim 0.95$, the accreted parcels that start farthest from periastron have initial velocities that produce retrograde trajectories. This effect extends towards periastron as $v_{\text{extra}} / v_{\text{per}}$ increases. At $v_{\text{extra}} / v_{\text{per}} = 1.00$ the parcel dropped from periastron falls to the accretor completely radially while all other accreted parcels follow retrograde trajectories. This is evident by the green dashed line in figure \ref{fig:vratio_v} crossing below 0, since retrograde impact velocities are taken as negative.

\begin{figure}[hbt!]
    \centering
    \includegraphics[width=\hsize]{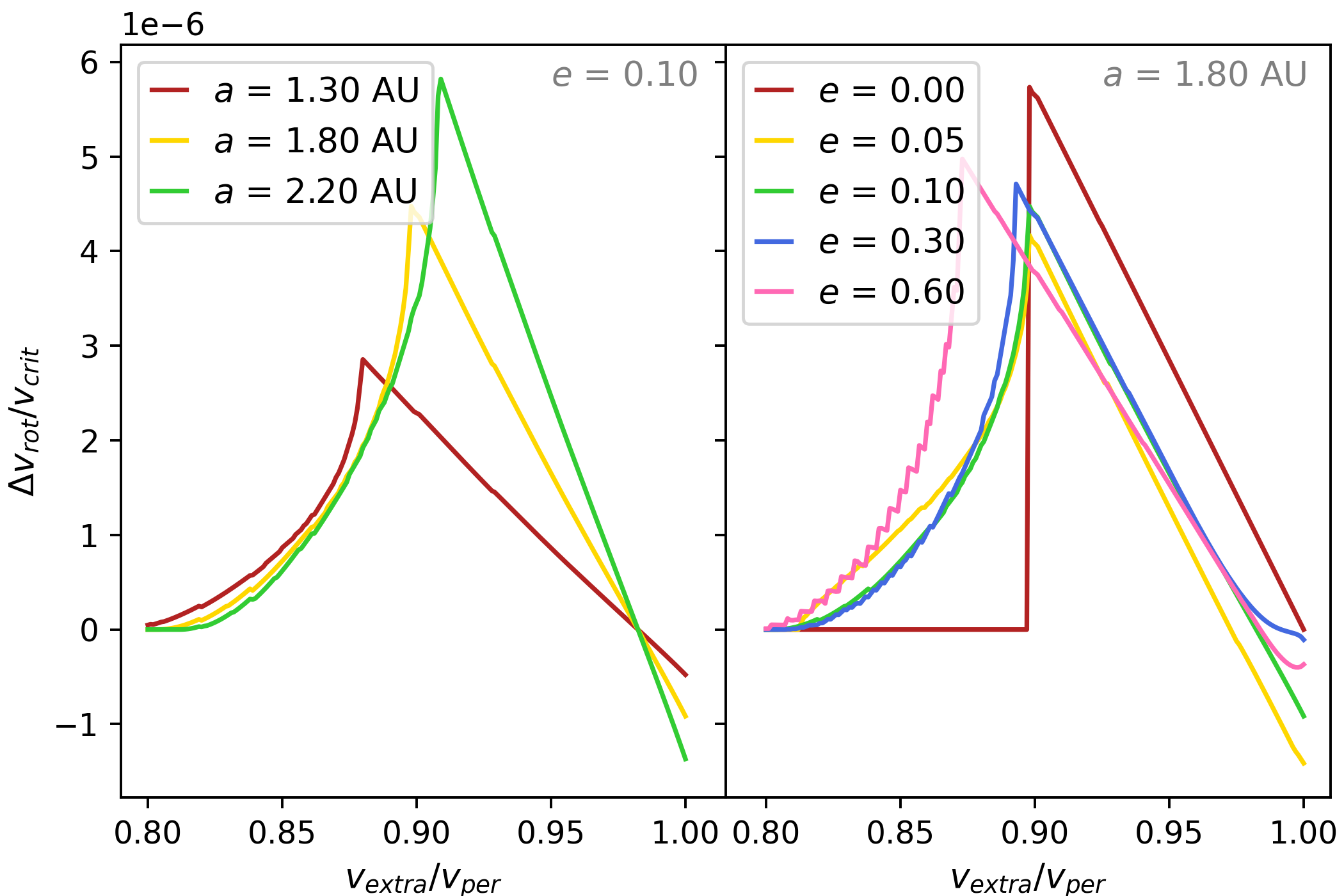}
    \caption{Spin-up of the accretor star in terms of its critical rotational velocity, as a function of the donor's rotational velocity $v_{\text{extra}} / v_{\text{per}}$. Left: All systems have $e = $ 0.10. The result is displayed for $a = $ 1.30, 1.80 \& 2.20 AU (red, yellow and green). Right: All systems have $a =  1.80$ AU and $e$ takes the values 0.00, 0.05, 0.10, 0.30, \& 0.60 (red, yellow, green, blue and pink).}
    \label{fig:spinup_v}
\end{figure}

For spin-up, the effects on the accretor as a function of $v_{\text{extra}} / v_{\text{per}}$ are displayed in figure \ref{fig:spinup_v}.

In the case of a circular orbit (red curve on right panel of figure \ref{fig:spinup_v}) direct impact occurs only for $v_{\text{extra}} / v_{\text{per}} \gtrsim 0.90$. This mirrors the behavior observed in the analysis of the circular case in section \ref{cont_a}. Here, the lowest $v_{\text{extra}} / v_{\text{per}}$ value where direct accretion occurs provides the most spin-up to the accreting star. This behavior is consistent for non-circular orbits, with the difference of a gradual increase towards the peak instead of the abrupt rise in spin-up.

\subsection{Conservativeness} \label{conservativeness}

We compute how conservative a system is by comparing the mass the donor looses $dm_{\text{don}}$ with the mass that directly impacts the accretor $dm_{\text{acc}}$ in one orbital period.

Within the orbit, the donor star sheds mass from its envelope where both conditions \ref{cond:of} \& \ref{cond:ang} are met simultaneously, while the accretor gains mass through direct accretion only when all three conditions are met. Under this premise, a fully conservative system is such that all mass leaving the donor directly impacts the accretor's surface. Here, parcels that do not impact the accretor directly will be counted as lost mass.

Any system where $|dm_{\text{acc}}/dm_{\text{don}}| < 1$ has a region in the orbit where the mass parcels lost by the donor are not directly accreted by the companion. If these parcels were to continue orbiting the accretor, a disk could be formed. Depending on the density of this disk in comparison to the density of the infalling parcels, a disk could prevent further direct accretion from occurring in a subsequent passage of the donor. Even if the parcels that miss the accretor might be able to continue orbiting the star and form a disk, this section's analysis focuses solely on the mass gained through direct accretion. Further discussion on the consequences of disk formation regarding the angular momentum transfer to the accretor will be discussed in a later section.

\subsubsection{Semi-major axis a}

\begin{figure}[hbt!]
    \centering
    \includegraphics[width=\hsize]{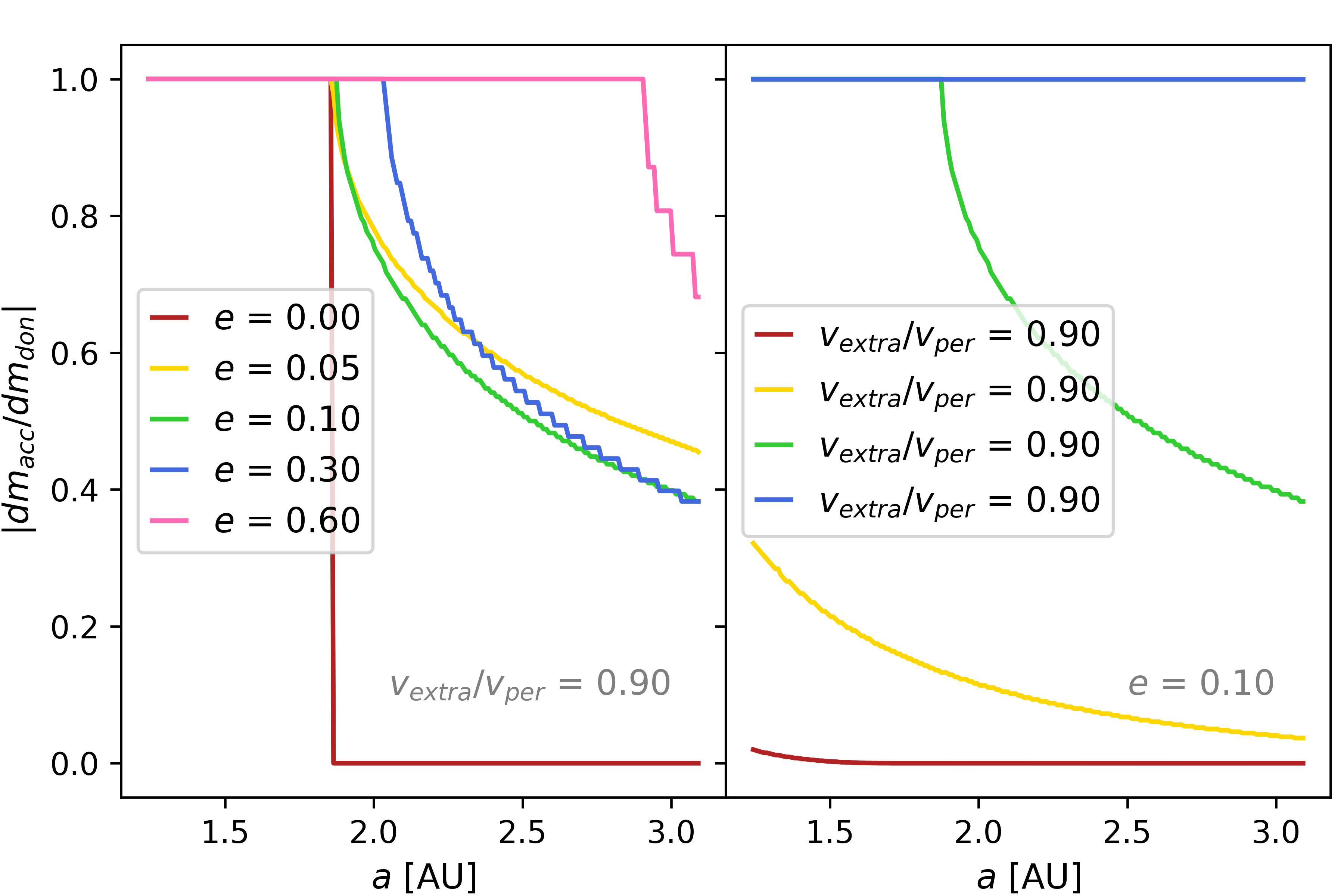}
    \caption{Conservativeness of the accretion event, as a function of the binary system's semi-major axis $a$. Left: All systems have $v_{\text{extra}} / v_{\text{per}} = $ 0.90. The result is displayed for $e = $ 0.00, 0.05, 0.10, 0.30 \& 0.60 (red, yellow, green, blue and pink). Right: All systems have $e =  0.10$ and $v_{\text{extra}} / v_{\text{per}}$ takes the values 0.80, 0.85, 0.90 \& 0.95 (red, yellow, green and blue).}
    \label{fig:cons_a}
\end{figure}

The conservativeness of these systems reflects the behavior seen in the spin-up curves in section \ref{cont_a}. By comparing those results with figure \ref{fig:cons_a}, we see that the systems with $a$ values lower than that of the peak in spin-up are completely conservative, while those with $a > a_{\text{peak}}$ present a decline in $|dm_{\text{acc}}/dm_{\text{don}}|$ due to condition \ref{cond:traj} forcing the accretion effective area to shrink away from periapsis. This is consistent with figure \ref{fig:a_diagram}.

The fully conservative nature of systems with $a \leq a_{\text{peak}}$ makes them the most effective at direct accretion. The fact that all the mass lost by the donor is gained by the accretor through direct accretion means there is no chance for a disk to form. Systems with $a > a_{\text{peak}}$, however, have a region close to periapsis where the donor's lost mass is not directly accreted.

\subsubsection{Eccentricity e}

In section \ref{cont_e} the tangential and radial contribution curves were divided into two cases: case 1 were systems that showed direct accretion for a circular orbit and appeared smooth (see the top panel of \ref{fig:vratio_e}); and case 2 were systems with no direct accretion at $e = 0$ and a noticeable peak somewhere in the distribution (see the bottom panel of figure \ref{fig:vratio_e}).
 
\begin{figure}[hbt!]
    \centering
    \includegraphics[width=\hsize]{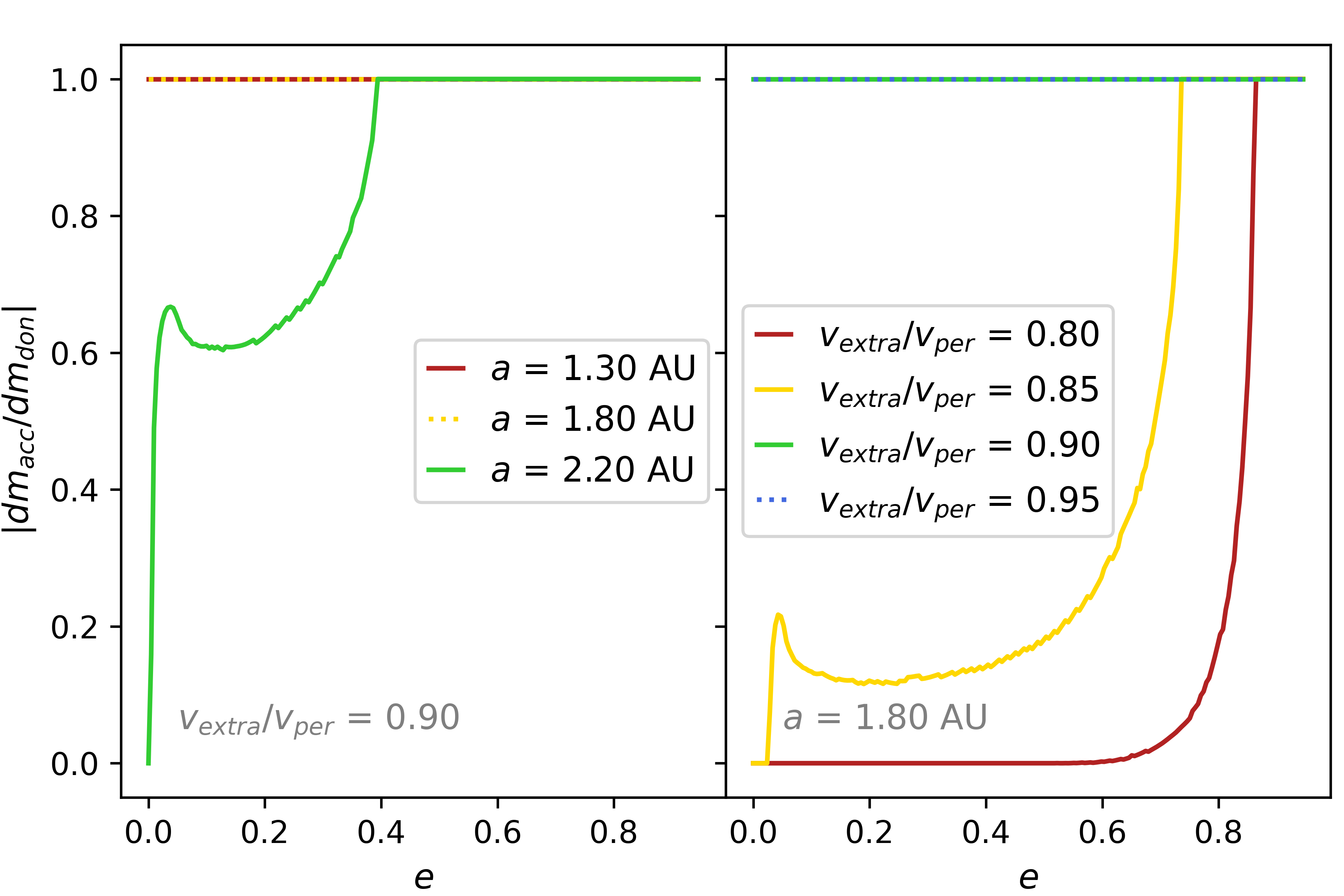}
    \caption{Conservativeness of the accretion event, as a function of the binary system's eccentricity $e$. Left: All systems have $v_{\text{extra}} / v_{\text{per}} = $ 0.90. The result is displayed for $a = $ 1.30, 1.80 \& 2.20 AU (red, yellow and green). Right: All systems have $a =  1.80$ AU and $v_{\text{extra}} / v_{\text{per}}$ takes the values 0.80, 0.85, 0.90, \& 0.95 (red, yellow, green and blue).}
    \label{fig:cons_e}
\end{figure}

Systems within case 1 are conservative through the whole range of $e$, as observed in the curves for $a = 1.30$ AU \& $a = 1.80$ AU in the left panel of figure \ref{fig:cons_e}, and the curves for $v_{\text{extra}} / v_{\text{per}} = 0.90$ \& $v_{\text{extra}} / v_{\text{per}} = 0.95$ in the right panel of the same figure.

We will use the curve for $a = 2.20$ AU (green) in the left panel of figure \ref{fig:cons_e} to analyze case 2 systems. The shape of this system's conservativeness curve is consistent with the behavior explored in the spin-up section for these same systems. None of the shed mass is directly accreted on a circular orbit due to condition \ref{cond:traj} failing to be met anywhere in the orbit. Condition \ref{cond:traj} is first fulfilled at apoapsis for very small $e$ and extends its domain towards periapsis as $e$ increases. Nonetheless, at $e \sim 0.05$ the donor is unable to fill its Roche lobe at the orbit's apoastron, making the region where condition \ref{cond:of} recede around periastron. The combination of these two factors is responsible for the shape of the curve at $0.00 < e \lesssim 0.4$. 

Around $e \sim 0.4$, the region of the orbit where condition \ref{cond:traj} is met reaches periapsis. This causes all instances of the orbit where the donor loses mass to produce particles with trajectories that impact the accretor (see stage "d" in figure \ref{fig:ediagram}).

\subsubsection{Donor's rotation}

\begin{figure}
    \centering
    \includegraphics[width=\hsize]{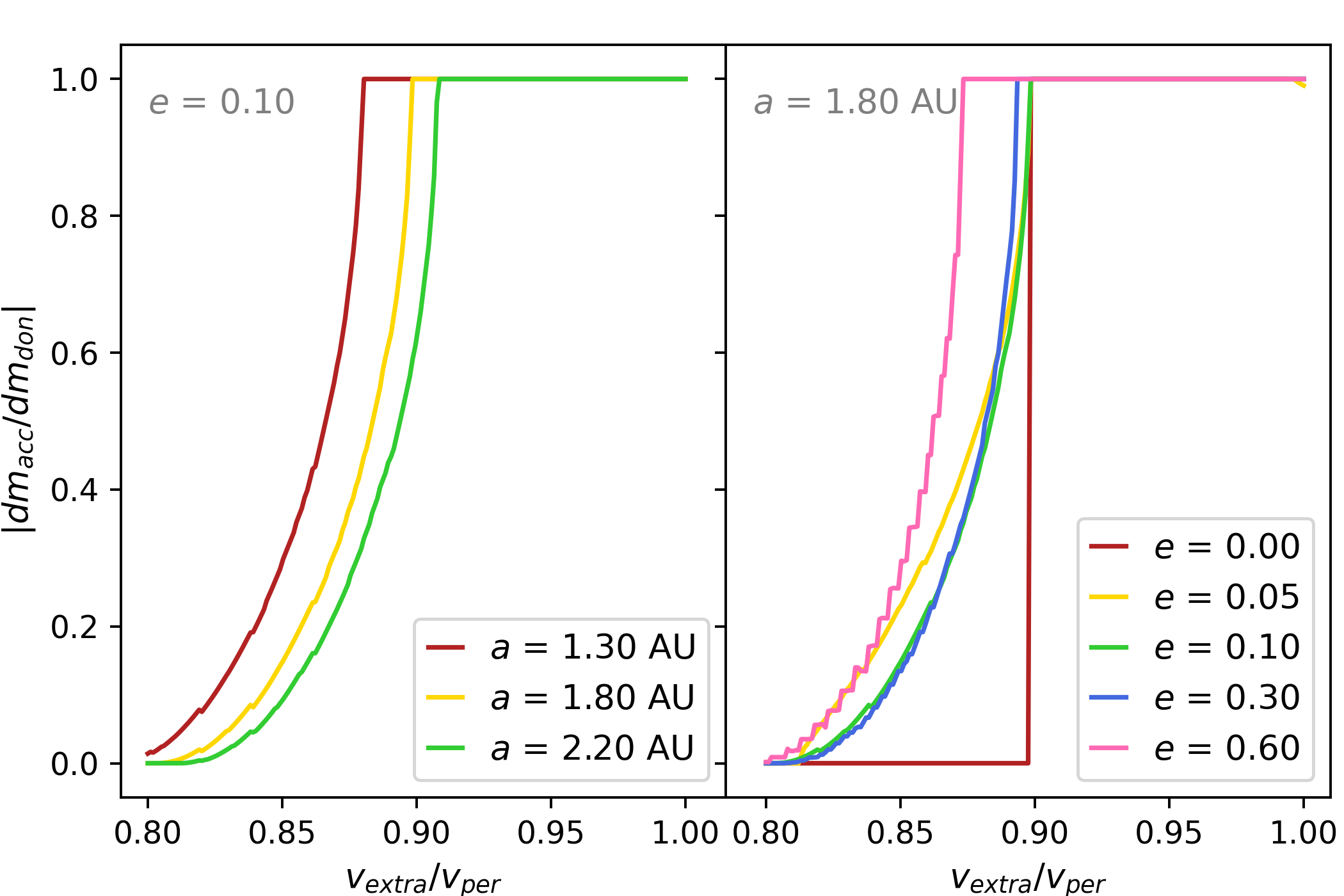}
    \caption{Conservativeness of the accretion event, as a function of the donor's rotational velocity $v_{\text{extra}} / v_{\text{per}}$. Left: All systems have $e = $ 0.10. The result is displayed for $a = $ 1.30, 1.80 \& 2.20 AU (red, yellow and green). Right: All systems have $a =  1.80$ AU and $e$ takes the values 0.00, 0.05, 0.10, 0.30, \& 0.60 (red, yellow, green, blue and pink).}
    \label{fig:cons_v}
\end{figure}

For all cases displayed in figure \ref{fig:cons_v}, conservativeness increases with $v_{\text{extra}} / v_{\text{per}}$ until reaching unity at the same donor rotation value where the peak in spin-up is achieved.

All systems with a donor rotation rate less than that of the peak present mass loss in the region near the periastron.

\subsection{Mass ratio q}

\begin{figure}
    \centering
    \includegraphics[width=\hsize]{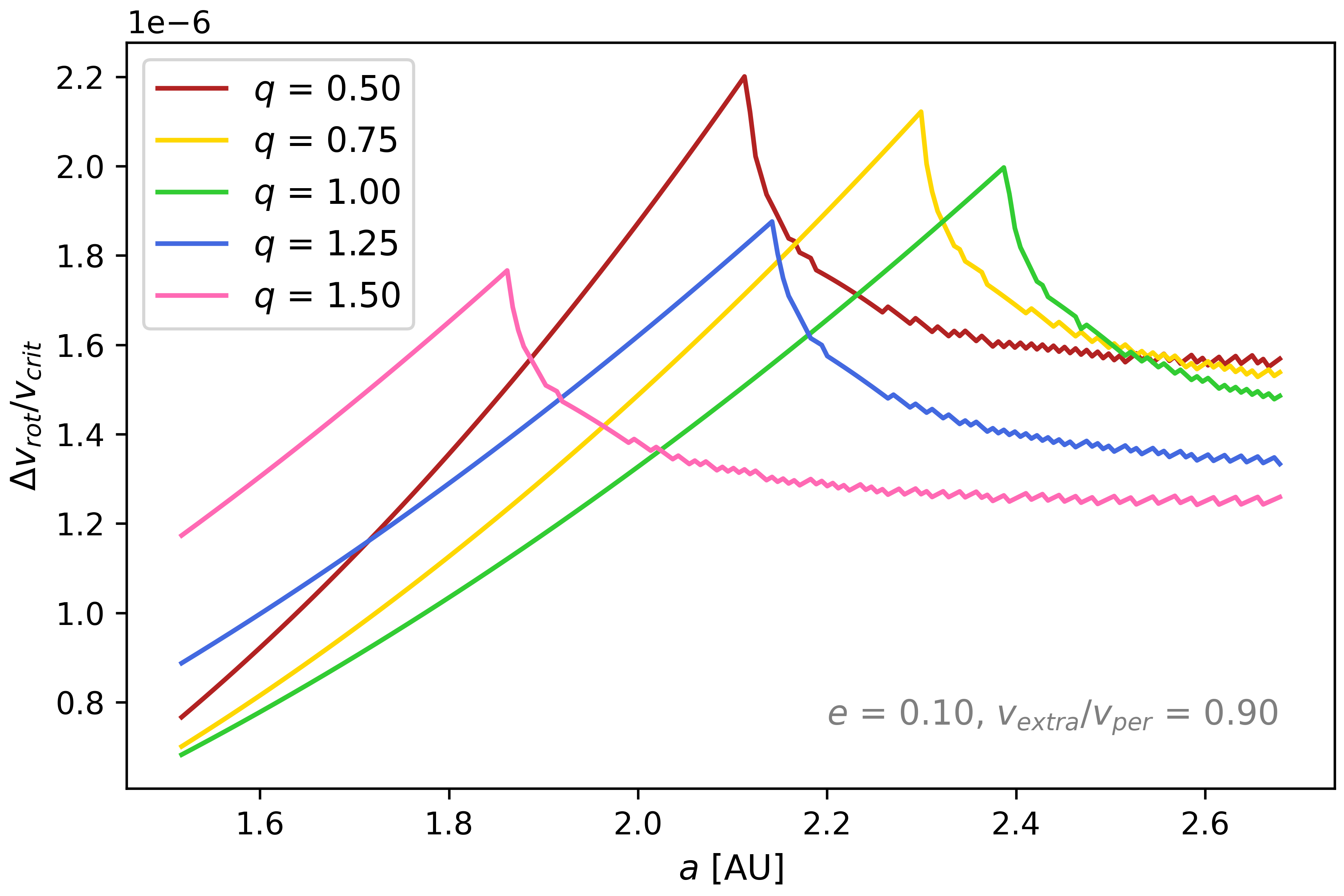}
    \caption{Alike figure \ref{fig:spinup_a} but for $q = $ 0.50, 0.75, 1.00, 1.25 \& 1.50 and constant $e =$ 0.10 and $v_{\text{extra}}/v_{\text{per}} = 0.90$.}
    \label{fig:q}
\end{figure}

Figure \ref{fig:q} shows the spin-up dependence on $a$ for different mass ratios $q = m_{\text{don}} / m_{\text{acc}}$. The accretor mass $m_{\text{acc}} = 1 M_{\odot}$ was kept constant and only the mass of the donor was changed to reflect different mass ratios. We can see that, when compared with figure \ref{fig:spinup_a}, the overall behavior of the curve is the same, but the position of the peak and its maximum spin-up value changes with $q$.

The value of $a_{\text{peak}}$ reaches its maximum value when the donor and accretor have the same mass and shifts towards smaller $a$ for both $q<1$ and $q>1$. Equation \ref{eq:L1} is dependent on the masses of both stars. Therefore, a change in the mass ratio also changes $r_{L1}$ (i.e. the starting point of a particle).

The maximum spin-up for all $q$ values is still of the order of $10^{-6} v_{\text{crit}}$, however, there is a subtle increase related to a decrease in $q$. For a donor mass of $m_{\text{don}} = 1.5 M_{\odot}$ the maximum spin-up is $\sim 1.8 \times 10^{-6} v_{\text{crit}}$ while a donor mass of $m_{\text{don}} = 0.5 M_{\odot}$ results in a spin-up of $\sim 2.2 \times 10^{-6} v_{\text{crit}}$. Although a $\sim 4 \times 10^{-7} v_{\text{crit}}$ might seem small, it is not negligible when compared with a spin-up effect of the order of $10^{-6} v_{\text{crit}}$. In future work, a wider exploration of the parameter space in regard to $q$ would help us understand more extreme cases. Nevertheless, we note that our model can properly account for different mass ratios.

\subsection{Timescale}

A spin-up of the order of $10^{-6} v_{\text{crit}}$ per period implies that, if the orbital parameters of the binary and the mass loss ratio of the donor stayed invariant, it would take $\sim 10^{6}$ periods ($0.92$ to $2.35$ Myr for $1.23$ AU $<a<$ $2.30$ AU) for the accretor to reach critical rotation. Yet, assuming our donor has a $0.53 M_{\odot}$ core and $\dot{M} = 10^{-6} M_{\odot}/$yr, it would only take $0.67$ Myr to completely shed its envelope. Thus, under constant orbital parameters and donor rotation rate, a star would not be able to spin-up to its critical rotation velocity under mass gain through direct accretion only. The timescales for changes to the orbital properties of the system will be discussed in the following section.

\section{Discussion} \label{Discussion}

\subsection{Synchronization} \label{synchronization}

For circular orbits, equation 33 of \citet{2011A&A...528A..48M} expresses the synchronization time
\begin{align*}
	\tau_{\text{syn}} = \frac{I (\beta_0^2 - 1) \omega_{\text{orb}}^2}{2 \dot{E}}
\end{align*}
, where $I$ is the moment of inertia and $\dot{E}$ the rate of energy dissipation. This timescale is dependent on the system's orbital separation "$a$" and the proportion of a star's rotation to the system's angular orbital velocity "$\beta_0$" (i.e. how synchronous the rotation is to the orbit, or $\omega_{\text{rot}} / \omega_{\text{orb}}$). A smaller $\beta_0$ results in a longer time, meaning that systems with a faster rotating donor would synchronize quicker than those that start closer to equilibrium. From figure 7 of their work, circular systems with a mass ratio $q = $ 1.25 ($m_{acc} = 4 M_{\odot}$ and $m_{don} = 5 M_{\odot}$), an orbital separation $a = $ 0.30 AU ($\log{(a / R_{\odot})} = 1.8$) and $v_{\text{extra}}/v_{\text{orb}} = $ 0.73 \& 1.70 ($\beta_0 = $ 1.2 \& 2.0 respectively) would have synchronization times of 100 Gyr and 25.12 Gyr respectively. It is important to note that the synchronization time is proportional to $a^6$, and since the minimum orbital separation value we explore ($a \sim 1.25$ AU) is greater than the maximum value shown in figure 7 from \citet{2011A&A...528A..48M}, we expect the synchronization times for our parameter space to be even greater.

\begin{figure}
	\centering
    \includegraphics[width=\hsize]{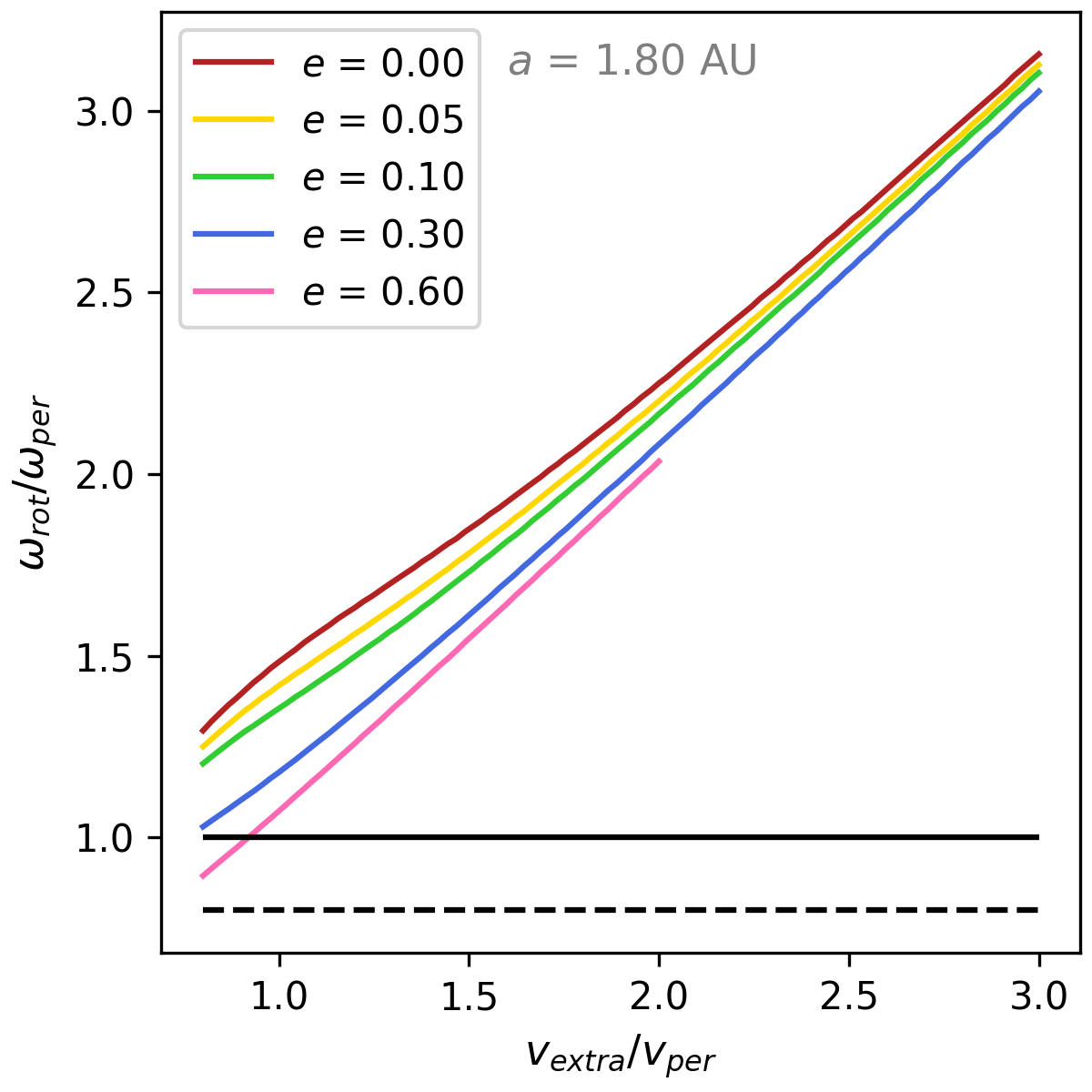}
    \caption{Akin to figure \ref{fig:synch} but for pseudo-synchronicity. The dashed line and solid black line represent the lower and upper limits for a pseudo-synchronous state (0.8 \& 1 respectively). Like in figure \ref{fig:alpha}, data for $e=$ 0.60 is not displayed beyond $v_{\text{extra}}/v_{\text{per}} =$ 2 because of numerical issues related to the resolution in the computation of $L1$.}
    \label{fig:synche}
\end{figure}

As briefly mentioned in section \ref{Validation} when comparing the circular case in our model to a planar restricted three-body approach, the minimum $v_{\text{extra}}/v_{\text{orb}}$ value needed to get a direct accretion event corresponds to a supersynchronous regime in both our model ($\omega_{\text{rot}} / \omega_{\text{orb}} \sim 1.3$) and the three-body model ($\omega_{\text{rot}} / \omega_{\text{orb}} \sim 2.2$). Figure \ref{fig:synche} shows a clearer comparison between the parameter $v_{\text{extra}}/v_{\text{per}}$ and synchronicity at periastron. In spite of that, we note that there is no need for the whole star to be spinning at such rates, but rather its outermost layers only. This could be achieved by dynamical tides affecting the surface. Nevertheless, further discussions about the mechanisms by which these rotational velocities could be obtained and maintained by the donor star are not part of this work, but will be included in future research.

\citet{1980A&A....92..167H} proved that a system can have up to two equilibrium states. These equilibrium state are characterized by: coplanarity between the equatorial planes of the stars and the orbital plane, circularity (i.e. $e=0.00$), and corotation between the stars' rotation periods and the orbital period. Supersynchronization deviates from these expected equilibrium states.

\begin{figure}
	\centering
    \includegraphics[width=\hsize]{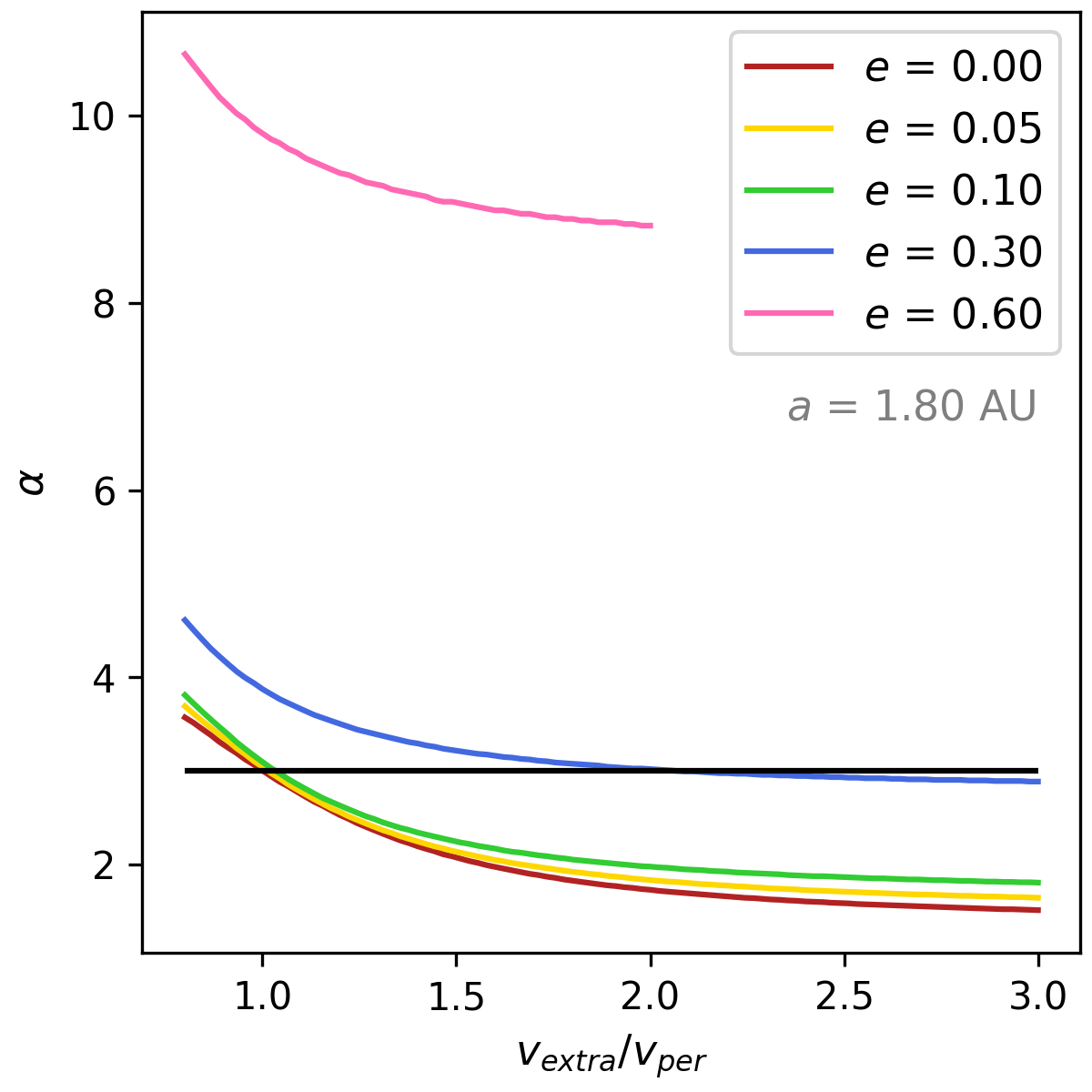}
    \caption{Ratio of orbital angular momentum to rotational angular momentum at the equilibrium state for a range of values of parameter $v_{\text{extra}}/v_{\text{per}}$. All systems have $a =$ 1.80 AU. Curves for $e =$ 0.00, 0.05, 0.10, 0.30 \& 0.60 (red, yellow, green, blue \& pink) are displayed. It is important to note that we only display the $e = 0.60$ data for $v_{\text{extra}}/v_{\text{per}} < 2$ due to numerical errors related to the computation of $L1$ that were affecting the results.}
    \label{fig:alpha}
\end{figure}

From the work of \citet{1973ApJ...180..307C} on tidal evolution and the stability of a two-body system upon perturbations, \citet{1980A&A....92..167H} found that the a state of equilibrium (coplanar, circular, corotating) is only stable when its ratio between orbital and rotational angular momentum is greater than 3. For elliptical orbits, \citet{Hut81} describes pseudo-synchronization as a state where rotation and orbital motion are near synchronous at periastron ($0.8 < \omega_{\text{rot}} / \omega_{\text{per}} \leq 1$). Pseudo-synchronization deviates from the unique equilibrium state since it forces the rotational period of the star to remain shorter than that of the orbit. Figure \ref{fig:alpha} shows the orbital to rotational angular momentum ratio $\alpha$ in contrast to our parameter $v_{\text{extra}}/v_{\text{per}}$. 

In \citet{Hut81}, the timescale for synchronization under perturbations from the stable equilibrium state is divided into three cases: (i) If $\alpha - 3 < \ll 1$ it takes the system a longer time to achieve synchronization, (ii) if $\alpha$ is of the order of 7 synchronization and changes in other parameters happen at similar timescales, and (iii) if $\alpha \gg 7$ the system tries to synchronize quickly but since the timescale for circularization of the orbit is much greater, the system enters a state of pseudo-synchronicity.

In figure \ref{fig:alpha} we see that for $v_{\text{extra}}/v_{\text{per}} < 1$, all systems fall above $\alpha = 3$. If we consider the three cases presented by \citet{Hut81}, we could expect a system with an eccentricity of $e = 0.60$ to pseudo-synchronize quicker than systems with lower eccentricities. 

\citet{2011A&A...528A..48M} also found timescales for pseudo-synchronization in the case of eccentric orbits:
\begin{align*}
	\tau_{\text{syn}} (\text{yr}) = 3.17 \times 10^{38} \frac{m_{\text{don}} r^2_{\text{don}} (\beta^2_0 - 1) (1 + e)}{\dot{E}_{\text{ave}} P (1 - e)^3}
\end{align*}
, where $P$ is the orbital period in days, $\dot{E}_{\text{ave}}$ is the orbit averaged energy dissipation rate, and $m_{\text{don}}$ and $r_{\text{don}}$ are in solar masses and solar radii respectively. From figure 8 of their work we can get the pseudo-synchronization timescales for systems with eccentricities $e =$ 0.00, 0.30, 0.70 \& 0.80 and $a = $ 1.23 \& 2.3 AU (minimum and maximum $a$ values we explore, respectively). These times can be found in table \ref{tab:syntimes}.

\begin{table}
	\caption{Synchronization ($e = 0.00$) and pseudo-synchronization ($e > 0.00$) times for systems with $a=$ 1.23 \& 2.30 and $e=$ 0.00, 0.30, 0.70 \& 0.80, extracted from figure 8 of \citet{2011A&A...528A..48M}. All systems have $\beta_0 = $ 1.2 ($v_{\text{extra}}/v_{\text{per}} =$ 0.73).}
	\label{tab:syntimes}
	\centering
	\begin{tabular}{c c c}
		\hline\hline
		$e$  & $\tau_{\text{syn}} (a = 1.23 \text{ AU})$ Gyr & $\tau_{\text{syn}} (a = 2.30 \text{ AU})$ Gyr \\
		\hline
		0.00 & $3.98 \times 10^5$                            & $3.98 \times 10^6$                            \\
		0.30 & $3.98 \times 10^2$                            & $1.58 \times 10^3$                            \\
		0.70 & $1.58 \times 10^2$                            & $3.16 \times 10^2$                            \\
		0.80 & $1 \times 10^2$                               & $2 \times 10^2$                               \\
		\hline
	\end{tabular}
\end{table}

Considering it only takes our donor 0.67 Myr to strip its envelope, we deem it reasonable to think of the donor's rotational velocity as invariable for our timescale.

\subsection{Evolution of orbital parameters}

Our results show the spin-up effect and mass gain of the accretor after a single orbital period, assuming the semi-major axis and eccentricity of the system remain constant during that span of time. Here, we compare this assumption with previously calculated timescales for changes in $a$ and $e$, caused by mass transfer.

\subsubsection{Semi-major axis a} \label{discussion_a}

In both conservative and non-conservative RLOF mass transfer, the mass change can either increase or decrease the orbital separation of the system \citep{2016ApJ...825...71D}.

In \citet{2010ApJ...724..546S}, timescales for the evolution of $a$ in systems undergoing direct accretion are calculated. The lower rightmost panel of figure 5 of their work covers our lower eccentricity systems ($0 \leq e \leq 0.10$) but for a slower donor rotation ($f_{1, i} = 0.90 \implies v_{\text{extra}}/v_{\text{per}} < 0.80$ for $e =$ 0.00, 0.05, 0.10 \& 0.30). Systems with mass ratios of the order of 1 would have a semi-major axis evolution timescale greater than 15 Gyr. Additionally, those systems where $q<1$ widen, while systems with $q>1$ experience a shrinkage of their orbit. Considering our timescale for envelope depletion is lesser than $1$ Myr, we believe it is fair to take $a$ as a constant in low eccentricity systems. 

To get a better understanding of the evolution of $a$ in eccentric orbits, we plan to expand this model to track the evolution of the semi-major axis and accretor's spin, as a consequence of the mass gain, in future work.

\subsubsection{Eccentricity e}

\citet{2005ApJ...620..970M} describe the greater orbital period $P'$ for which a binary with the most frequent initial eccentricity in a population circularizes within the lifetimes of the stars. In other words, binary systems with initial periods shorter than $P'$ are expected to have circularized by the age of the population. Their method was tested on eight late-type binary populations with ages ranging from $\sim 3$ Myr to $\sim 10$ Gyr. The value of $P'$ increases with the population's age but, for all the discussed cases, does not grow far beyond $10$ days. The domain of semi-major axis values we explore in this investigation ($\sim 1.25 - 2.30$ AU) result in orbital periods ranging from $\sim 340$ days to $\sim 860$ days, much greater than the tidal circularization period of $\sim 10$ days for populations as old as $10$ Gyr. In consequence, we do not expect the eccentricity of our systems to significantly change from their initial values after mass transfer.

The three cases from \citet{Hut81} we discussed in section \ref{synchronization} can also give us some insight on the circularization times: When $\alpha$ is of the order of (greater than) 7, the circularization time will be similar to (greater than) the synchronization time. Since the synchronization times (see table \ref{tab:syntimes}) are many orders of magnitude greater than the time needed for the donor to deplete its envelope, circularization should not be a concern for these cases either. On the other hand, the case where $\alpha - 3 \ll 1$ results in a quicker circularization. Figure 6 from \citet{2010ApJ...724..546S} presents timescales for the evolution of $e$. The lower rightmost panel of their figure (like in section \ref{discussion_a}) covers a parameter space closest to ours. There, they obtained an eccentricity evolution timescale lower than 1 Gyr for systems with initial eccentricities $0.00 < e < 0.10$.

Overall, the circularization timescales for $e >$ 0.10 are larger than our depletion time by many orders of magnitudes and choosing to keep $e$ constant should not present any problem. However, the exact changes in eccentricity for systems with low initial eccentricities ($e \leq$ 0.10) and $v_{\text{extra}}/v_{\text{per}} \sim$ 1 remain unconstrained. In addition to tracking the evolution of the semi-major axis, we too plan to track the evolution of a system's eccentricity during a mass gain event in our future work.

\subsection{Penetration depth} \label{penetrationdepth}

As stated in section \ref{Results}, in this analysis the accretor is assumed to behave as a solid sphere. In reality, a stream impacting a star would penetrate through the surface and deposit its momentum in an inner layer. The depth at which the momentum is deposited and how it is redistributed inside the accretor will, in consequence, affect its observational properties (e.g. color, temperature, hot spots).

A MS star like our accretor has a density gradient and a pressure gradient with their maximum values at the star's center. We expect an infalling parcel of mass, with an associated density and pressure, to penetrate beyond the surface to a layer where either the star's density or pressure is equal or greater than that of the parcel.

In future work, we will expand this model to include a more realistic description of the accretor's internal structure so we can better analyze how the transferred momentum is deposited into each layer of the stellar structure. 

\subsection{Disk formation}

In this work's analysis we have solely focused on the mass that forms a stream and directly impacts the surface of the accretor. Any mass that did not hit the accretor upon approaching it was counted as lost mass, when in reality it could still be gravitationally bound to the accretor. If many of these parcels were to continue orbiting the accretor, an accretion disk could form around it. The presence of a disk around the accretor could force any infalling mass to interact with it before reaching the accretor's surface, thus preventing further direct accretion from occurring.

Additionally, during disk accretion, the inner edge of the disk follows a circular Keplerian orbit at the surface of the accretor, impacting the accretor fully tangentially. If the orbital parameters of the system were to remain constant, this would maximize spin-up.

Given that the addition of disk accretion would not only affect the total mass gain but also the observed spin-up effect, we expect our results to reflect a realistic outcome only for systems where direct accretion is completely conservative, or those where the mass that is not directly accreted becomes unbound from the system.

In future work, we aim to discern between the mass that forms a disk and the mass that is lost from the system.

\section{Conclusions} \label{Conclusions}

We presented a novel analytical model to approximate the spin-up effect an accreting star suffers due to direct accretion in a binary system. We explore a range of values of the system's semi-major axis $a$ (1.23 to 2.30 AU), its eccentricity $e$ (0.00 to 0.95), and the donor's linear rotational velocity, in terms of its linear orbital velocity at periastron, $v_{\text{extra}} / v_{\text{per}}$ (0.80 to 1.00).

Our model neglects the effect of the donor's gravitational potential and instead approaches the problem as a particle-accretor two-body system. Yet we demonstrate that the two-body approximation can easily be corrected to reproduce the three-body approximation in our example, making the approximations accurate (see section \ref{Validation}).

In regard to the gained momentum and the spin-up effect suffered by the accretor star, we find:
\begin{enumerate}
	\item Systems with small $a$ maximize the momentum deposited into the thermal energy of the accretor, while wider systems maximize the energy deposited towards the accretor's spin-up.
	
	\item For a fixed eccentricity and donor rotational velocity, the spin-up effect is maximized when the semi-major axis equals a value $a_{\text{peak}}$ (see equation \ref{eq:apeak}), for which the mass lost from periastron arrives to the accretor's surface tangentially. The smaller $a$ is in comparison to $a_{\text{peak}}$, the smaller the contribution to spin-up is. This implies that direct accretion in close binaries contributes more to the thermal energy of the star than its rotational energy.
	
	\item Systems with smaller eccentricities contribute more to the rotation of the accretor than its thermal energy. The opposite is true for more eccentric binaries.	
	
	\item For a fixed semi-major axis and donor rotational velocity, the dependence on eccentricity can follow one of two trends: If all conditions for direct accretion are satisfied for $e = 0.00$ (case 1), a circular orbit is the most effective at spinning up the accretor. If a system does not present direct accretion for $e=0.00$ (case 2), a system with the same initial conditions other than eccentricity (i.e. mass ratio, semi-major axis and donor rotation rate) is the most effective at spinning up the accretor when its eccentricity is the smallest value for which direct accretion occurs at periastron ($e_{\text{peak}}$).
	
	\item Systems whose donors have slower rotation rates maximize the momentum transferred to the rotation of the donor, while systems with greater $v_{\text{extra}} / v_{\text{per}}$ contribute the most to the thermal energy of the accretor.
	
	\item For a fixed semi-major axis and eccentricity, the accretor is spun up the most for the smallest $v_{\text{extra}} / v_{\text{per}}$ value for which direct accretion takes place at periastron. For greater $v_{\text{extra}} / v_{\text{per}}$ ($\leq 1.00$), direct accretion becomes more radial and therefore the contribution to rotational (thermal) energy is minimized (maximized).

	\item The maximum spin-up effect we obtain for direct accretion is of the order of $10^{-6}$ times the accretor's critical rotational velocity, per orbital period. This means that, if the initial orbital parameters were to remain constant, it would take $10^6$ periods to spin up the accretor to its critical rotation.
\end{enumerate}

In regard to mass transfer efficiency for direct accretion:
\begin{enumerate}
	\item Assuming $a$, $e$, $v_{\text{extra}} / v_{\text{per}}$ and the donor mass loss ratio $\dot{M} = 10^{-6} M_{\odot} /$yr remain constant, it takes the donor between $7.28 \times 10^{5}$ ($a =$ 1.23 AU) and $2.85 \times 10^{5}$ ($a =$ 2.30 AU) periods (0.67 Myr) to completely shed its envelope. Under the same premise, it would take the accretor around $10^{6}$ periods ($0.92$ to $2.35$ Myr for $1.23$ AU $<a<$ $2.30$ AU) to reach critical rotation. This implies that, under this premise, the accretor would be able to accrete more than just $10\%$ of its initial mass while avoiding break-up velocity.
	
	\item For a fixed eccentricity and donor rotational velocity, if $a \leq a_{\text{peak}}$, all the mass lost by the donor is directly accreted by the accretor. When the opposite is true ($a \geq a_{\text{peak}}$) direct accretion only occurs at the farthest point from periastron where the donor is still able to overfill its Roche lobe ($f > 0.00$) and mass is lost around periastron.
	
	\item For a fixed semi-major axis and donor rotational velocity, if all conditions for direct accretion are satisfied for $e = 0.00$, mass transfer is completely conservative for any $e$ (case 1). For case 2, mass transfer is only conservative for $e > e_{\text{peak}}$.
	
	\item For a fixed semi-major axis and eccentricity, mass transfer is totally conservative for any $v_{\text{extra}} / v_{\text{per}}$ value greater than that of the peak in spin-up. When $v_{\text{extra}} / v_{\text{per}}$ is smaller than that critical value, the mass lost by the donor near periastron is not directly accreted.
\end{enumerate}

We also note that:
\begin{enumerate}
	\item The maximum spin-up per orbit for systems with $q = 0.50$ and $q = 1.50$ differ in $4\times 10^{-8} v_{\text{crit}}$. Although small, this difference is not negligible when compared to a maximum spin-up of the order of $10^{-7} v_{\text{crit}}$. An wider exploration of the parameter space for $q$ will aid us in describing the spin-up's dependence on the binary's mass ratio.
		
	\item The minimum $v_{\text{extra}} / v_{\text{per}}$ value for direct accretion ($\sim 0.80$ in our model and $\sim 1.8$ for the three body approach in section \ref{Validation}) imply supersynchronization as a requirement for direct accretion. Although we do not discuss the mechanisms that would cause the donor star to achieve such a rotation rate, we do note that only the outermost layers of the donor are required to be supersynchronous to the orbit.
\end{enumerate}

In addition to the more thoroughly studied $a$ and $e$ effects on accretion, the rotation of the donor seems to also affect the angle of impact of the accreted mass. In consequence, this influences the angular momentum transferred to the accretor's rotation. The evolution of the donor's rotation, either through its stellar evolution or through tidal effects, can change the way a system accretes.

This work provides an analytical model for direct accretion that can quickly provide limits on the parameter space for the semi-major axis, eccentricity and donor rotation rates.

In the next work,we plan to track the orbital evolution of the system as a response of mass transfer through our model's prescription. This will allow us to directly compare with the timescales for $a$ and $e$ found in the literature. We also seek to refine this model by exploring how the depth within the star at which the accreted mass is deposited can influence its properties. Lastly, we aim to discern between the lost mass and the parcels that stay bound to the system so we can better describe the whole accretion event in the case an accretion disk forms.

\begin{acknowledgements}
      We acknowledge financial support from Millenium Nucleus NCN19{\_}058 \& NCN2023{\_}002 (TITANs). We are thankful to Silvia Toonen and Alonso Herrera-Urquieta for providing the SeBa data used in this work. NWCL gratefully acknowledges the generous support of a Fondecyt General grant 1230082, as well as support from Millenium Nucleus NCN19{\_}058 (TITANs).
\end{acknowledgements}

%
%

\bibliographystyle{aa} 
\bibliography{refs}

\begin{appendix}

\section{Linear eccentricity of a parcel's orbit}

In section \ref{Model}, the eccentricity of a parcel's trajectory towards the accretor is defined as 
\begin{align*}
	e_p = \frac{c_p}{a_p}
\end{align*}
, where $c_p$ and $a_p$ are the linear eccentricity and semi-major axis of the parcel's orbit. In this appendix, we derive $c_p$ from the geometry of the parcel's trajectory.

\begin{figure}[hbt!]
    \centering
    \includegraphics[width=\hsize]{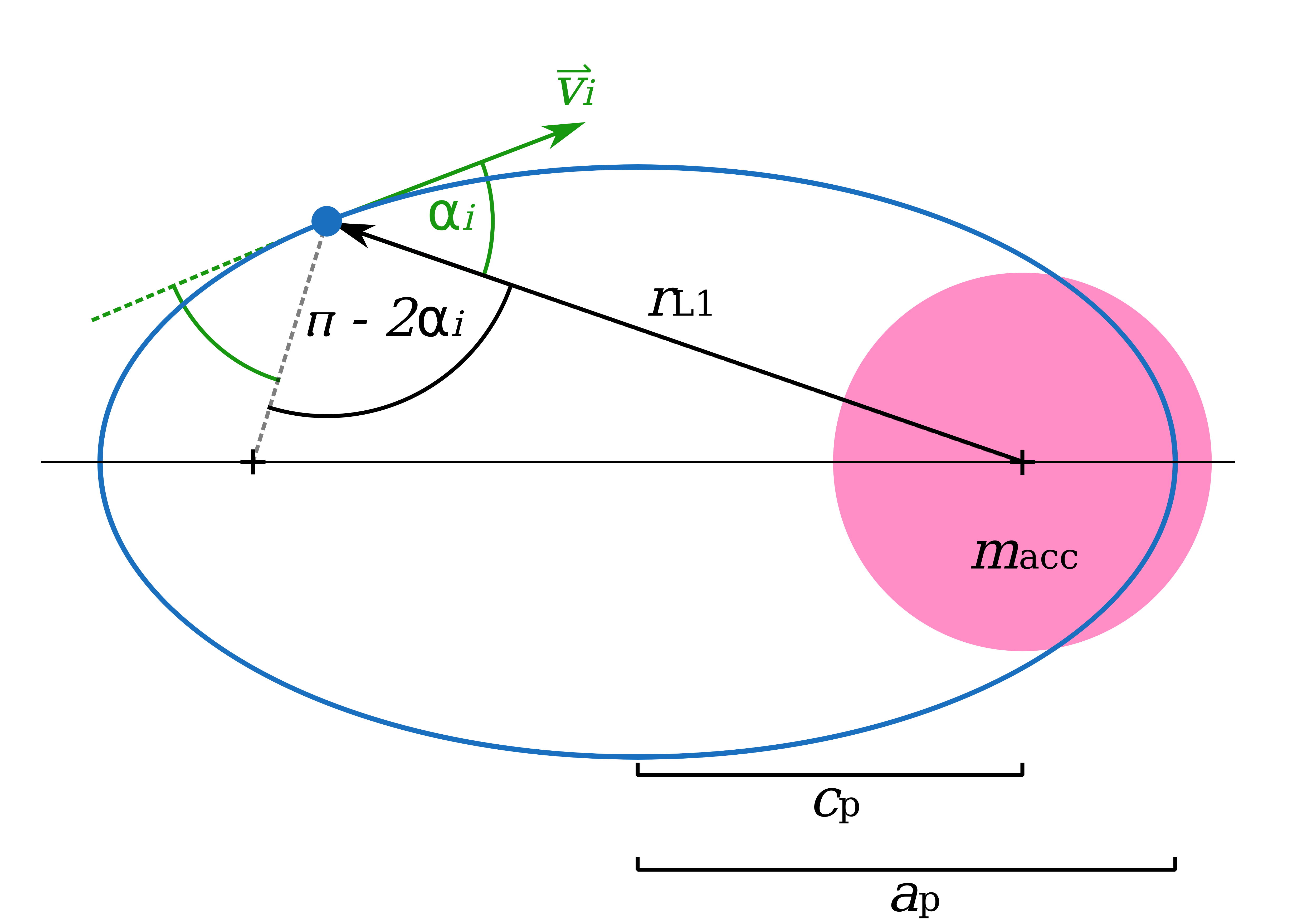}
    \caption{Diagram.}
    \label{fig:parcelorbit}
\end{figure}

For any point of the ellipse, the sum of its distances to the focal points is constant and equal to $2a_p$. Then, the length of the segment that joins the parcel to the focal point that does not align with the accretor is $2a_p - r_{L1}$.

As shown in figure \ref{fig:parcelorbit}, the angle between the two segments that join the parcel to the focal points can be written in terms of the parcel's velocity angle $\alpha_i$, obtaining
\begin{align*}
	\pi - 2 \alpha_i 
\end{align*}
Knowing the length of two sides of the triangle and the angle between them, we can use the law of cosines to get the missing side
\begin{align*}
	(2c_p)^2 = r_{L1}^2 + (2 a_p - r_{L1})^2 - 2 r_{L1} (2 a_p - r_{L1}) \cos{(\pi - 2\alpha_i)}
\end{align*}
. Finally, we arrive to equation \ref{eq:cp}
\begin{align*}
	c_p = \frac{1}{2} \sqrt{ r_{L1}^2 + (2 a_p - r_{L1})^2 - 2 r_{L1} (2 a_p - r_{L1}) \cos{(\pi - 2\alpha_i)} }
\end{align*}

\section{Limits on $v_{\text{extra}}/v_{\text{per}}$} \label{apx:vfr_limit}

\begin{figure}[hbt!]
    \centering
    \includegraphics[width=\hsize]{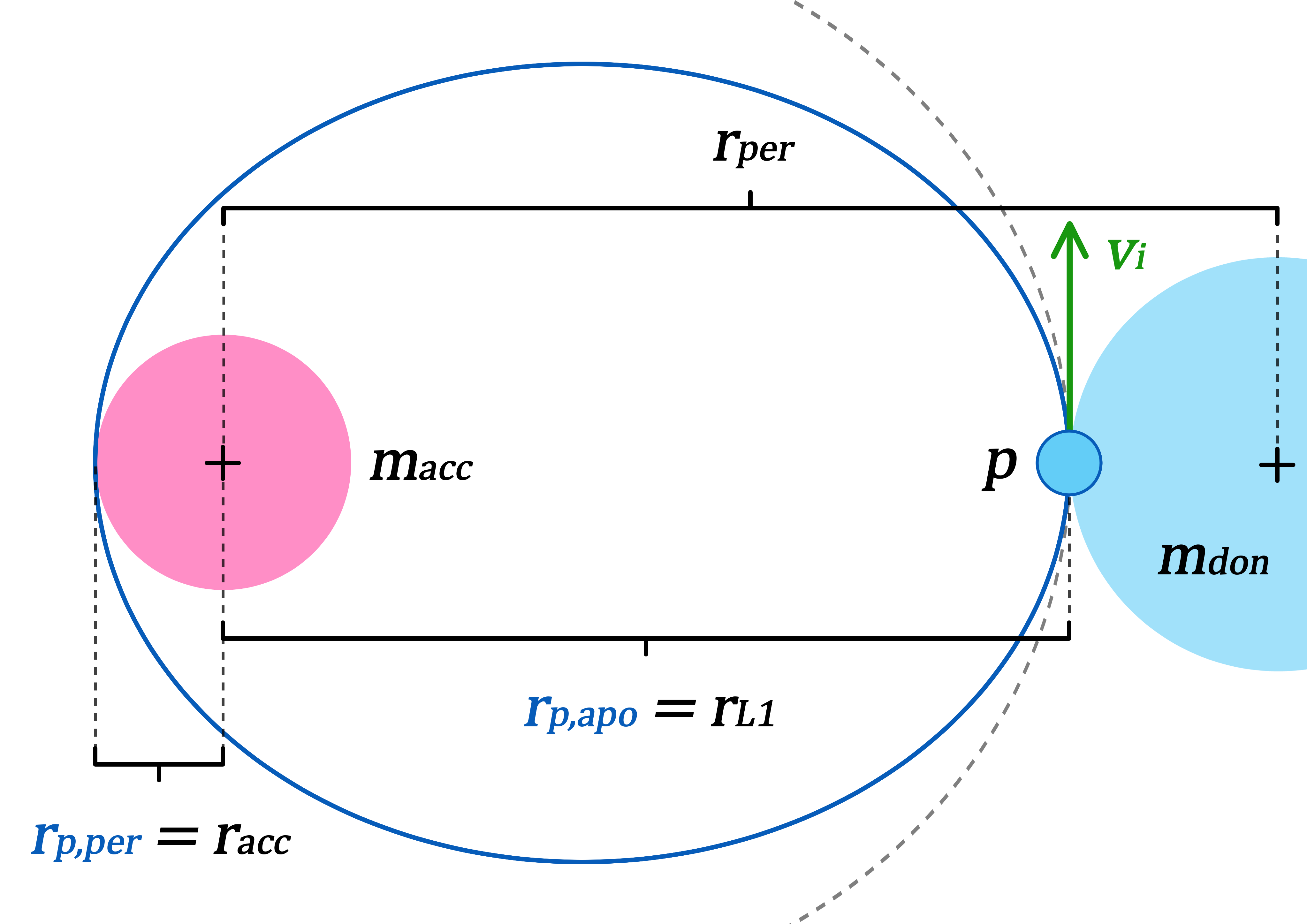}
    \caption{Diagram showing the initial configuration of a parcel being dropped from periastron. The accretor and donor are drawn in pink and light blue respectively. The parcel is denoted with the letter $p$ and its orbit around the accretor is drawn in dark blue.}
    \label{fig:dvlim}
\end{figure}

At periastron, a particle is shed from a donor star's envelope and falls from a distance $r_{L1}$, onto a central body of mass $m_{acc}$, with an initial velocity $v_i$. This particle will move on an orbit of eccentricity $e_p$ and semi-major axis $a_p$. If $v_i$ is perpendicular to the direction of the initial position of the particle $\Vec{r}_{L1}$, this starting position must be either the periapsis or apoapsis of its orbit. Taking assumption \ref{I} into consideration, an initial velocity equal to
\begin{align}
    v_i = \sqrt{\frac{G m_{acc}}{r_{L1}}}
\end{align}
will result in a circular orbit. For lower velocities, the parcel's orbit will shrink, decreasing its semi-major axis and become more eccentric, making the initial position at $r_{L1}$ the apoapsis of the orbit:
\begin{align} \label{eq:rapo}
    r_{p, apo} = r_{L1}
\end{align}
We then locate the center of mass of the central body which corresponds to the foci closest to periapsis. If the central body were to have a radius $r_{acc}$, the particle would impact onto the central body's surface when
\begin{align} \label{eq:rper}
    r_{p, per} \leq r_{acc}
\end{align}
where $r_{p, per}$ is the particle's periapsis distance. In turn, given equation \ref{eq:rapo} and the highest possible value for $r_{p, per}$ in expression \ref{eq:rper}, the greatest length of the semi-major axis of the particle's orbit so that it still impacts the central body is
\begin{align}
    a_p = \frac{r_{acc} + r_{L1}}{2}
\end{align}
Replacing this within the vis-viva equation (\ref{eq:visviva}), we obtain the minimum initial velocity required for an orbit that impacts the central body
\begin{align} \label{eq:vilim}
    v_i = \sqrt{\frac{2Gm_{acc}}{r_{L1} + r_{L1}^2 / r_{acc}}}
\end{align}
This configuration is exemplified by figure \ref{fig:dvlim}.

From equation \ref{eq:vi}, we know that at periastron, where the orbital and extra velocity are perpendicular to each other, the particle's initial velocity will be given by
\begin{align} \label{eq:viper}
    v_i = v_{per} - v_{extra}
\end{align}
where $v_{per}$ is the velocity of the donor star around the accretor star at periastron. This velocity can be obtained through the vis-viva equation (\ref{eq:visviva})
\begin{align} \label{eq:vper}
    v_{per} = \sqrt{G(m_{acc} + m_{don}) \left( \frac{2}{r_{per}} - \frac{1}{a} \right)}
\end{align}
Dividing equation \ref{eq:viper} by $v_{per}$ and replacing \ref{eq:vilim} and \ref{eq:viper}, we obtain an expression for the lowest value the synchronicity parameter $v_{extra} / v_{per}$ can take to ensure direct accretion occurs
\begin{align} \label{eq:fracv}
    \frac{v_{extra}}{v_{per}} = 1 - \sqrt{\frac{2 m_{acc}}{(r_{L1} + r_{L1}^2 / r_{acc})(m_{acc} + m_{don}) (2/r_{per} - 1/a)}}
\end{align}

\begin{align} \label{eq:vi_in_vfr}
    v_i = v_{per} \left( 1 - \frac{v_{extra}}{v_{per}} \right)
\end{align}

\begin{align}
    \frac{G m_{don}}{(r_{per} - r_{L1})^2} - \frac{G m_{acc}}{r_{L1}^2} + \frac{2 G m_{acc}}{r_{L1}^2 + r_{L1}^3 / r_{acc}} = 0
\end{align}

\section{Initial angle} \label{apx:initial_angle}

\begin{figure}[hbt!]
    \centering
    \includegraphics[width=\hsize]{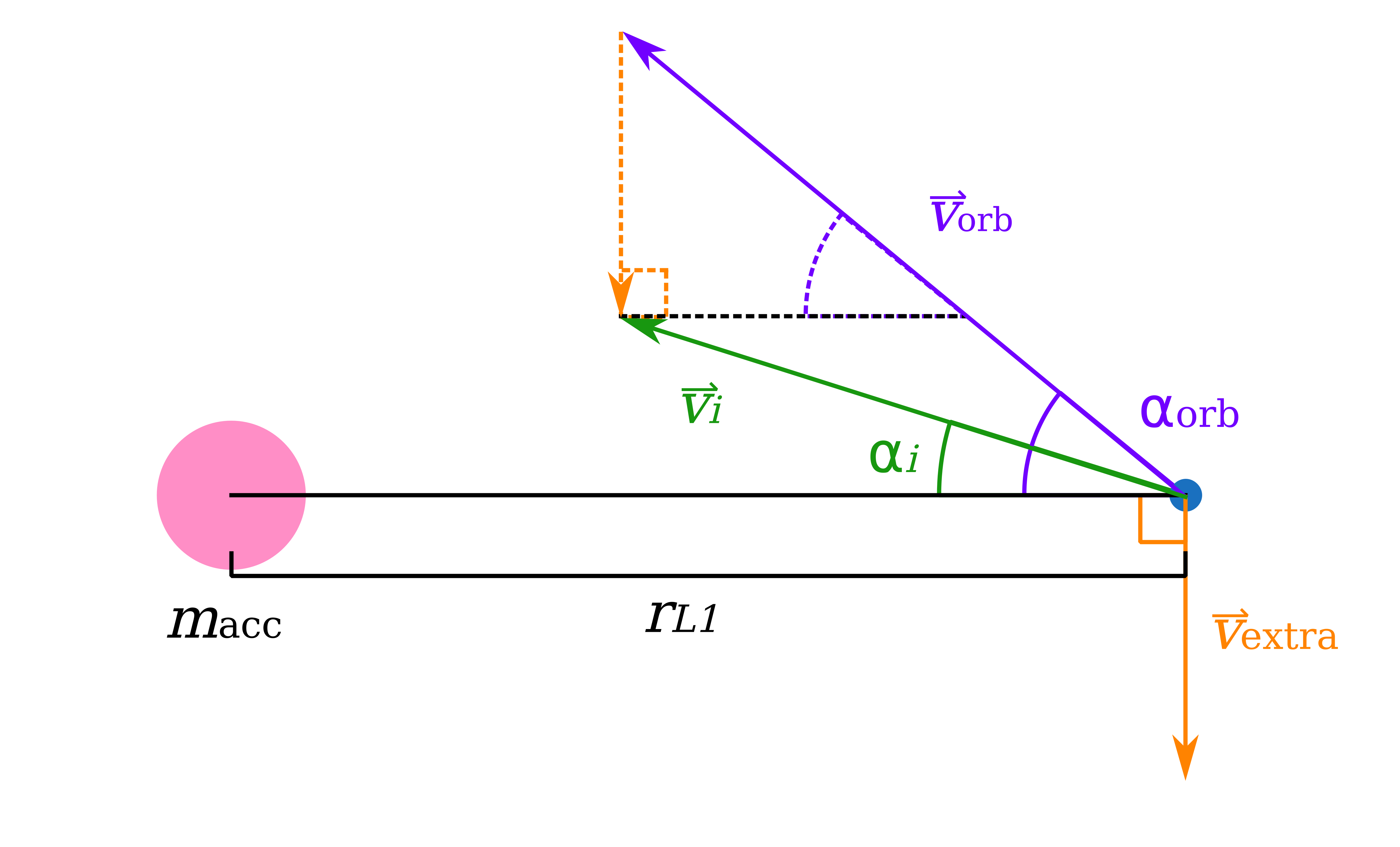}
    \caption{Diagram of the angles with respect to the radial direction of velocities $\vec{v}_i$ (green), $\vec{v}_{\text{orb}}$ (purple) and $\vec{v}_{\text{extra}}$ (orange). Velocity vectors and their associated angles are drawn in solid lines while projections are drawn in dashed lines.}
    \label{fig:initial_angle}
\end{figure}

At the starting point, a parcel has an initial velocity $\vec{v}_i$
\begin{align}
	\vec{v}_i = \vec{v}_{\text{orb}} + \vec{v}_{\text{extra}}
\end{align}
The initial angle with respect to the radial direction depends on the magnitude and direction of $\vec{v}_{\text{orb}}$ and $\vec{v}_{\text{extra}}$. Following figure \ref{fig:initial_angle}, we use the law of sines to write
\begin{align*}
	\frac{\sin{\left( \alpha_{\text{orb}} + \alpha_i \right)}}{v_{\text{extra}}} = \frac{\sin{\left( \frac{\pi}{2} - \alpha_{\text{orb}} \right)}}{v_{i}}
\end{align*}
Then, applying the arcsine function and solving for $\alpha_i$ we are left with
\begin{align}
	\alpha_i = \alpha_{\text{orb}} + \arcsin{\left[ \frac{v_{\text{extra}}}{v_i} \sin{\left( \frac{\pi}{2} - \alpha_{\text{orb}} \right)} \right]}
\end{align}

\clearpage

\section{Best fit polynomial coefficients}

\begin{table}[hb]
\caption{Polynomial coefficients for the best fit to the impact angle-donor rotation velocity relation. The coefficients of the polynomial fit are shown in decreasing order, from left to right (i.e. P6 to P0).}\label{tab:coefficients}
\centering
\begin{tabular}{ccccccccc} 
\hline\hline             
$q$ & $a$ [AU] & P6 & P5 & P4 & P3 & P2 & P1 & P0\\
\hline
	\multicolumn{9}{c}{\it This work} \\
\hline
  0.5 & 1.3 & 7.45$\times 10^{6}$ & -4.23$\times 10^{7}$ & 1.00$\times 10^{8}$ & -1.26$\times 10^{8}$ & 8.97$\times 10^{7}$ & -3.39$\times 10^{7}$ & 5.35$\times 10^{6}$\\
  0.5 & 1.8 & 1.47$\times 10^{7}$ & -8.41$\times 10^{7}$ & 2.01$\times 10^{8}$ & -2.56$\times 10^{8}$ & 1.84$\times 10^{8}$ & -7.03$\times 10^{7}$ & 1.12$\times 10^{7}$\\
  0.5 & 2.2 & 3.45$\times 10^{7}$ & -1.99$\times 10^{8}$ & 4.77$\times 10^{8}$ & -6.11$\times 10^{8}$ & 4.39$\times 10^{8}$ & -1.69$\times 10^{8}$ & 2.70$\times 10^{7}$\\
  1.0 & 1.3 & 7.23$\times 10^{6}$ & -4.10$\times 10^{7}$ & 9.70$\times 10^{7}$ & -1.22$\times 10^{8}$ & 8.67$\times 10^{7}$ & -3.28$\times 10^{7}$ & 5.16$\times 10^{6}$\\
  1.0 & 1.8 & 1.09$\times 10^{7}$ & -6.23$\times 10^{7}$ & 1.49$\times 10^{8}$ & -1.90$\times 10^{8}$ & 1.36$\times 10^{8}$ & -5.21$\times 10^{7}$ & 8.29$\times 10^{6}$\\
  1.0 & 2.2 & 2.46$\times 10^{7}$ & -1.42$\times 10^{8}$ & 3.40$\times 10^{8}$ & -4.35$\times 10^{8}$ & 3.13$\times 10^{8}$ & -1.20$\times 10^{8}$ & 1.92$\times 10^{7}$\\
  1.5 & 1.3 & 6.69$\times 10^{6}$ & -3.80$\times 10^{7}$ & 9.00$\times 10^{7}$ & -1.14$\times 10^{8}$ & 8.06$\times 10^{7}$ & -3.05$\times 10^{7}$ & 4.81$\times 10^{6}$\\
  1.5 & 1.8 & 9.37$\times 10^{6}$ & -5.38$\times 10^{7}$ & 1.29$\times 10^{8}$ & -1.64$\times 10^{8}$ & 1.18$\times 10^{8}$ & -4.51$\times 10^{7}$ & 7.19$\times 10^{6}$\\
  1.5 & 2.2 & 2.93$\times 10^{7}$ & -1.69$\times 10^{8}$ & 4.05$\times 10^{8}$ & -5.18$\times 10^{8}$ & 3.73$\times 10^{8}$ & -1.43$\times 10^{8}$ & 2.29$\times 10^{7}$\\
\hline
  \multicolumn{9}{c}{\it Three-body} \\
\hline
  0.5 & 1.3 & 1.91$\times 10^{7}$ & -1.41$\times 10^{8}$ & 4.33$\times 10^{8}$ & -7.08$\times 10^{8}$ & 6.52$\times 10^{8}$ & -3.20$\times 10^{8}$ & 6.55$\times 10^{7}$\\
  0.5 & 1.8 & -2.23$\times 10^{8}$ & 1.66$\times 10^{9}$ & -5.14$\times 10^{9}$ & 8.49$\times 10^{9}$ & -7.88$\times 10^{9}$ & 3.91$\times 10^{9}$ & -8.06$\times 10^{8}$\\
  0.5 & 2.2 & 3.54$\times 10^{6}$ & -1.78$\times 10^{7}$ & 2.80$\times 10^{7}$ & 4.44$\times 10^{4}$ & -4.45$\times 10^{7}$ & 4.48$\times 10^{7}$ & -1.41$\times 10^{7}$\\
  1.0 & 1.3 & 7.26$\times 10^{4}$ & -7.63$\times 10^{5}$ & 3.34$\times 10^{6}$ & -7.78$\times 10^{6}$ & 1.02$\times 10^{7}$ & -7.15$\times 10^{6}$ & 2.09$\times 10^{6}$\\
  1.0 & 1.8 & 8.48$\times 10^{5}$ & -8.94$\times 10^{6}$ & 3.93$\times 10^{7}$ & -9.22$\times 10^{7}$ & 1.22$\times 10^{8}$ & -8.55$\times 10^{7}$ & 2.50$\times 10^{7}$\\
  1.0 & 2.2 & 6.32$\times 10^{5}$ & -6.72$\times 10^{6}$ & 2.98$\times 10^{7}$ & -7.03$\times 10^{7}$ & 9.34$\times 10^{7}$ & -6.61$\times 10^{7}$ & 1.95$\times 10^{7}$\\
  1.5 & 1.3 & 4.72$\times 10^{3}$ & -6.37$\times 10^{4}$ & 3.58$\times 10^{5}$ & -1.07$\times 10^{6}$ & 1.81$\times 10^{6}$ & -1.62$\times 10^{6}$ & 6.08$\times 10^{5}$\\
  1.5 & 1.8 & 7.06$\times 10^{3}$ & -9.64$\times 10^{4}$ & 5.49$\times 10^{5}$ & -1.67$\times 10^{6}$ & 2.84$\times 10^{6}$ & -2.59$\times 10^{6}$ & 9.80$\times 10^{5}$\\
  1.5 & 2.2 & 3.13$\times 10^{4}$ & -4.28$\times 10^{5}$ & 2.44$\times 10^{6}$ & -7.40$\times 10^{6}$ & 1.26$\times 10^{7}$ & -1.15$\times 10^{7}$ & 4.36$\times 10^{6}$\\
\hline
\end{tabular}
\end{table}

\end{appendix}






   
  



\end{document}